\documentclass[12pt]{elsarticle}

\usepackage[letterpaper,margin=1.0in]{geometry}
\usepackage{amsmath,amsfonts,amssymb}
\usepackage[toc,page]{appendix}
\usepackage[T1]{fontenc}
\usepackage{newtxtext,newtxmath}
\usepackage{libertine}
\usepackage{microtype}
\DisableLigatures[f]{encoding = *, family = * }
\usepackage{textcomp}
\usepackage{graphicx,subfigure,epstopdf}
\usepackage[dvipsnames]{xcolor}
\usepackage{setspace}
\usepackage{fancyhdr}
\usepackage{enumerate,etaremune}
\usepackage{tabularx,multirow}
\usepackage{framed}
\usepackage{hhline}
\usepackage[pdftex,unicode,bookmarks=false]{hyperref}
\usepackage{url}
\hypersetup{
  colorlinks,
  citecolor=MidnightBlue, linkcolor=MidnightBlue, urlcolor=MidnightBlue
}
\usepackage{tikz}
\usetikzlibrary{calc}
\usepackage{algorithm}
\usepackage[noend]{algpseudocode}
\usepackage{mathtools}

\usepackage{lineno}

\newcommand{\tensor}[1]{\ensuremath{\boldsymbol{#1}}}

\DeclareMathOperator{\tr}{tr}

\biboptions{numbers,sort&compress,square,comma}
\AtBeginDocument{\hypersetup{citecolor=MidnightBlue,linkcolor=MidnightBlue,urlcolor=MidnightBlue}}

\makeatletter
\def\ps@pprintTitle{%
 \let\@oddhead\@empty
 \let\@evenhead\@empty
 \def\@oddfoot{}%
 \let\@evenfoot\@oddfoot}
\makeatother

\begin{document}

\begin{frontmatter}

\title{A single Long Short-Term Memory network for enhancing prediction of path-dependent plasticity with material heterogeneity and anisotropy}

\author[Institution_1]{Ehsan Motevali Haghighi}
\ead{motevale@mcmaster.ca}

\author[Institution_1]{SeonHong Na \corref{corr}}
\ead{nas1@mcmaster.ca}

\cortext[corr]{Corresponding Author}

\address[Institution_1]{Department of Civil Engineering, McMaster University, Hamilton, ON L8S 4L8, Canada}

\journal{.}

\begin{abstract}
This study presents applicability of conventional deep recurrent neural networks (RNN)
to predict path-dependent plasticity associated with material heterogeneity and anisotropy. 
Although the architecture of RNN possess inductive biases toward information over time, 
it is a still challenging to to learn the path-dependent material behavior as a function of loading path
considering the change from elastic to elasto-plastic regimes. 
Our attempt is to develop a simple machine-learning based model that can replicate 
elastoplastic behaviors considering material heterogeneity and anisotropy. 
The basic Long-Short Term Memory Unit (LSTM) is adopted
for the modeling of plasticity in the two-dimensional space by enhancing the inductive bias 
toward the past information through manipulating input variables.  
Our results find that a single LSTM based model can capture the J2 plasticity responses
under both monotonic and arbitrary loading paths provided the material heterogeneity.
The proposed neural network architecture is then used to model elasto-plastic responses 
of a two-dimensional transversely anisotropic material 
associated with computational homogenization (FE$^2$).
It is also found that a single LSTM model can be used to accurately and effectively capture 
the path-dependent responses of heterogeneous and anisotropic microstructures
under arbitrary mechanical loading conditions.
\end{abstract}


\end{frontmatter}


\section{Introduction}
\label{sec:intro}
Natural and artificial processes, diverse interactions among living and non-living things,
are easily found as historical dependent, referred to as path-dependent or time-dependent phenomena. 
The path-dependence is defined as a behavior in which the next step of an action, 
 in a non-temporal sense, is affected by the previous steps. 
For instance, the auto-complete feature of the Google search engine demonstrates the path-dependent behavior, 
where the past search history guides the next search.
In this case, there is no time concept 
because the machine only uses the past information without considering the time or duration of typing words. 
On the other hand, there is a time-dependent behavior
in which the next time step depends on the previous time steps.
For example, the time of releasing the company's new product has significant consequences 
on the company's stock \cite{chen2015lstm}. 
The complex mixture behavior of path-dependency and time-dependency is observed from the transportation system
when both stopping time and selecting the route of one driver will affect 
all drivers on the systems \cite{javadinasr2021deep}.  

In the context of computational mechanics, the prediction of a material's behavior 
is one of the evident examples of path-dependent and time-dependent behaviors.
For instance, permanent deformation of the material
due to the dislocation of crystal structures indicates the path-dependent behavior, 
which is usually captured by constitutive laws - crystal plasticity \cite{anand1996computational,clayton2010nonlinear,na2018computational}. 
An example of materials' time dependence can be found from creep,
which is expressed by a function of time in general \cite{urai1986weakening,chan1998recovery}.

Despite the advancement of constitutive laws for complicated mechanical behaviors of solids,
capturing their anisotropy or heterogeneity is still a challenging task. 
One of the main reasons for this challenge arises from the highly diverse 
heterogeneous and anisotropic system, which sometimes require various internal parameters
associated with ad-hoc constitutive laws \cite{na2017effects}.
To remedy this issue, multiscale methods have been proposed 
\cite{di2012optimizing,dvorak1976axisymmetric,moulinec1994fast},
which consider heterogeneity and anisotropy by explicitly modeling microstructures. 
However, the computational cost of running the micro-scale calculations interacted with the macro-scale computation
limits the potential applicability of the multiscale approach. 
One of the emerging methods to reduce that computational cost for the multiscale simulation 
is using a surrogate model or a data-drive model of microstructural behaviors 
by leveraging Artificial Intelligence \cite{mozaffar2019deep,ghavamian2019accelerating}.

Initially, the artificial neural networks was used to develop constitutive models 
by using experimental data sets  \cite{ghaboussi1991knowledge,ghaboussi1998autoprogressive,pernot1999application}. 
Instead of capturing generalized loading responses of solids, 
these approaches focused on how to train neural networks 
based on the constitutive information, which is usually constrained by testings,
and how to overcome this constraint. 
Interestingly, it was reported that the path-dependent behavior of homogeneous microstructures 
cannot be predicted via Multilayered Perceptron Network (MLP) \cite{homel2019machine}.  
On the other hand, it was also demonstrated to predict the homogeneous micro-structural responses 
via Artificial Neural Network (ANN) with the correction of input data 
- including the averaged past strain \cite{palau2012neural}. 
Recently, a Gated Recurrent Network (GRU) is implemented to identify plasticity-constitutive laws
of general materials, which predicts stress tensor by feeding strain tensors 
associated with adding descriptors for several microstructures \cite{mozaffar2019deep}.

Various efforts have also been made to apply the deep neural networks
for multiscale modeling \cite{wang2018multiscale,wang2019updated}.
For example, a Long-Short Term Memory Unit (LSTM) was adopted to predict behaviors of homogeneous microstructures,
in which the TensorFlow's auto-differentiation for multiscale simulations \cite{ghavamian2019accelerating}. 
The path-dependent behavior of homogeneous microstructures was also investigated
as a surrogate model, which was implemented to the macro-level 
using the Long-Short Term Memory Unit (LSTM) \cite{capuano2019smart}. 
Recently, a reinforcement learning method for hyper-parameter tuning was developed \cite{fuchs2021dnn2}, 
which captured the path-dependent behavior of a specific micro-structure. 

Several recent studies also focused on adopting deep neural networks for multiscale simulations of heterogeneous solids
\cite{xu2020data,liu2019deep,im2021surrogate}.
As an example, a computational framework to establish a data-driven constitutive model for heterogeneous path-dependent composites 
has been implemented to predict the stress-strain relationships via the principal values \cite{ge2021computational},
in which adopted separate data-driven models were adopted for elastic and plastic parts, respectively. 
A recurrent neural network-accelerated multi-scale model for elastoplastic heterogeneous materials 
subjected to random cyclic and non-proportional loading paths 
was investigated by considering a single microstructure \cite{wu2020recurrent}. 
Within the small-strain regime, both linear and non-linear elastic responses of heterogeneous microstructures were captured  
by feeding probabilistic descriptors as an input \cite{haghighi2021multifeatured}. 
However, less attention has been paid to directly identify the path-dependent relationship between the stress and strain tensors 
for diverse material heterogeneity under generalized loading conditions. 

In this study, a single data-driven framework is proposed to predict heterogeneous path-dependent responses of solids
by leveraging the Long-Short Term Memory unit (LSTM), 
which is capable of capturing both elastic and elastoplastic increments. 
Previously, the basic LSTM was reported not to capture the elastoplastic responses
due to its lacking consideration of coupled energy conservation-dissipation mechanisms \cite{hoedt2021mc}.
In the proposed framework, however, this issue is resolved by directly feeding the past averaged history of strains as input,
associated with the recurrent neural network architecture. 

This paper is divided into three parts to investigate the performance of a single LSTM network architecture 
in capturing the path-dependent behavior of various microstructures. 
In Part 1, verification of a FE$^2$ homogenization framework is conducted
using a benchmark problem presented by \citet{peric2011micro}. 
This framework is then used to generate homogenized responses of transversely isotropic microstructures
for collecting their path-dependent responses for training (Part 3). 
Next, capability of the conventional LSTM  approach is investigated
through the J2 plasticity in Part 2.  
The proposed LSTM is tested with the constitutive law considering material heterogeneity. 
Model parameters, including elastic constants, hardening modulus, and yield stress, are randomly generated 
to account for the material heterogeneity.
14,000  sets of heterogeneous material properties are considered, 
and randomly generated loading paths, including monotonic loading-unloading, 
are applied to identify their path-dependent responses. 
In Part 3, finally, applicability of the basic LSTM is investigated
in terms of extracting and learning the path-dependent anisotropic responses of microstructures. 
The anisotropy of microstructures, transversely isotropic,  is explicitly defined by multiple horizontal layers
with alternating elastic and elastoplastic constitutive laws. 
Different material properties are selected to consider heterogeneous constitutive information,
where geometrical descriptors are adopted to describe explicitly configured transversely isotropic microstructures. 
Our results demonstrate the capability of a conventional LSTM in predicting heterogeneous and path-dependent behavior 
without deterioration of static data (descriptors like material properties and geometrical descriptors) 
when it is fed align with dynamic data (strain tensor) to the network during a sequence.

The organization of this paper is as follows. 
In Section 2, the objectives of this study are presented by addressing two approaches
associated with mechanical responses of materials. 
In Section 3, the framework for J2 plasticity constitutive law and 
homogenization techniques are presented. 
In Section 4, a brief review of the deep neural networks and the architecture of the long-short term memory unit are depicted. 
In Section 5, the design of the experiment (DOE) is presented, which includes the generation 
of loading path, heterogeneous path-dependent response, and anisotropic microstructures. 
Finally, three parts of investigation are demonstrated 
to validate the homogenization framework and test the deep LSTM for capturing path-dependent behavior in Section 6.  
The following notations and symbols are used throughout: bold-face letters 
denote tensors and vectors; the symbol ``$\cdot$'' denotes an inner product of
 two vectors (e.g., $\tensor{a}\cdot\tensor{b}=a_{i}b_{i}$), 
:or a single contraction of adjacent indices of two tensors
 (e.g., $\tensor{c}\cdot\tensor{d}=c_{ij}d_{jk}$); the symbol ``$:$'' denotes an
  inner product of two second-order tensors (e.g., $\tensor{c}:\tensor{d}=c_{ij}d_{ij}$).
Following the standard mechanics sign convention, stress is positive in tension and pressure
 is positive in compression.

\section{Problem statements}
The primary objective of this study is to demonstrate deep recurrent neural networks
to reproduce heterogeneous and anisotropic path-dependent behaviors 
either from constitutive laws or from microstructural homogenization. 
To begin with, the J2 plasticity constitutive law is considered as a reference model 
to generate a database of heterogeneous path-dependent behavior 
by randomly selecting the model's parameter. 
Then the FE$^2$ homogenization method is implemented 
to collect homogenized responses of randomly generated layered 2D domain
as a reference for mimicking anisotropy of microstructures. 
We may address the following problems to investigate the capability of deep neural networks 
for heterogeneous and anisotropic path-dependent behaviors: 

\begin{itemize}
\item \textbf{Problem I}: 
Stress tensor responses against strain-driven loading under random, uniaxial, and biaxial conditions.
A deep neural network ($f_h$-h stands heterogeneity) is designed 
to predict 2D Cauchy stress tensors ($\tensor{\sigma_t}$) along paths 
by feeding a sequence of strain tensors($\tensor{\epsilon_t}$), 
model parameters ($\Upsilon =[\lambda, \mu, \sigma_{y0}, H]$) of heterogeneous system, 
and averaged strain ($\frac{\sum_{0}^{t}\tensor{\epsilon_t}}{t}$):
\begin{equation}
f_h: (\tensor{\epsilon_t}, \Upsilon, \frac{\sum_{0}^{t}\tensor{\epsilon_t}}{t} ) \rightarrow (\tensor{\sigma_t})
\end{equation}  
\item \textbf{Problem II}:
Stress tensor responses against strain-driven loading under random, uniaxial, and biaxial conditions. 
A deep neural network ($f_a$-a stands anisotropy) is designed 
to predict the 2D homogenized stress tenors ($\tensor{\sigma_t}$) along paths 
by feeding a sequence of strain tensors ($\tensor{\epsilon_t}$), 
microstructural descriptors ($\Upsilon$ =[Microstructural Descriptor]) of an anisotropic system, 
and the averaged strain ($\frac{\sum_{0}^{t}\tensor{\epsilon_t}}{t}$):
\begin{equation}
f_a : (\tensor{\epsilon_t}, \Upsilon, \frac{\sum_{0}^{t}\tensor{\epsilon_t}}{t}) \rightarrow (\tensor{\sigma_t})
\end{equation}  
\end{itemize}

\section{Plasticity constitutive model and computational homogenization}
\subsection{Isotropic hardening J2 plasticity}
For completeness, this section reiterates the isotropic hardening J2 plasticity model 
with its kinematics and yield criterion \cite{borja2013plasticity}.
The additive decomposition of stress tensor into volumetric and deviatoric parts gives, 
\begin{equation}
\tensor{\sigma}={p}\tensor{1}+\tensor{s},
\end{equation}
where $\tensor{\sigma}$ is Cauchy stress tensor, ${p}=tr(\tensor{\sigma})/3$ denotes the mean normal stress,
$\tensor{1}$ is the second ranked identity tensor, and $\tensor{s}$ is the deviatoric stress tensor
satisfying the condition $tr(\tensor{s})=0$, in which $\tr$ is the trace operator. 
Similarly, the additive decomposition of infinitesimal strain tensor $\tensor{\epsilon}$ can be given as,
\begin{equation}
\tensor{\epsilon}=\frac{1}{3}{\epsilon_v}\tensor{1}+\tensor{e},
\end{equation}
where ${\epsilon}_v=\tr(\tensor{\epsilon})$ indicates the volumetric strain,
and $\tensor{e}$ the deviatoric strain tensor.
For the isotropic linearly elastic regime, the elastic constitutive equations are:
\begin{equation}
{p}=K {\epsilon}_v, \\\ \tensor{s}=2\mu \tensor{e},
\end{equation}
where $K$ and $\mu$ are the elastic bulk and shear moduli, respectively. 
Therefore, the overall relationship between stress and strain in the isotropic elastic case can be obtained as, 
\begin{equation}
\tensor{\sigma} = K{\epsilon}_v\tensor{1}+2\mu\tensor{e} = \tensor{C}^{e} :\tensor{ \epsilon},
\end{equation}
where
\begin{equation}
\tensor{C}^e=K\tensor{1}\otimes\tensor{1}+2\mu(\tensor{I}\ -\frac{1}{3}\tensor{1}\otimes\tensor{1}),
\end{equation}
is the rank-four tensor of elastic moduli. The J2 yield function can be obtained as,
\begin{equation}
f(\tensor{\sigma},\kappa)=\sqrt{2J_2}-\kappa\le0,
\end{equation}
where $J_2$ is the second invariant of the deviatoric stress tensor $\tensor{s}$ and defined as,  
\begin{equation}
J_2= \frac{1}{2} s_{ij} s_{ij}.
\end{equation}
Therefore, the elastic region can be closed as
\begin{equation}
\overline{E}=\left\{\left(\tensor{\sigma},\kappa\right)\in\mathbb{S}\times\mathbb{R}^1\ |\ f\left(\tensor{\sigma},\kappa\right)\le0\right\},
\end{equation}
where $\mathbb{S}$ is the space of linear, second-order symmetric tensor, and $f(\tensor{\sigma},\kappa)=0$ defines the yield surface for $J_2$ plasticity. Defining a variable $\kappa$, one can determine the behavior is softening or hardening by satisfying consistency equation as, 
\begin{equation}
\frac{\partial{f}}{\partial\sigma}\\: \dot{\sigma}-H\dot{\lambda} = 0 \ \ \ \text{with} \ \ \
H=-\frac{\partial{f}}{\partial{\kappa}}\left(\frac{\dot{\kappa}}{\dot{\lambda}}\right),
\end{equation}
and the evolution of $\kappa$ with plastic strain must be of the form:,
\begin{equation}
\kappa=\frac{2}{3}H^{'}\lambda+\kappa_0,
\end{equation}
where $H'$ is the plastic modulus, and $\kappa_0$ is the reference value of the $\kappa$ when $\lambda=0$.
Please refer to \citet{borja2013plasticity} for more details.

\subsection{Homogenization method}
\label{sec : FEM2}
To localize the macro strain path on the boundary of the representative volume element (RVE), 
we use the linear displacement boundary condition on the micro-level \cite{miehe2002computational}. 
Based on this method, the deformation boundary constraints in terms of the macro strain 
$(\tensor{ \varepsilon}_M)$ can be obtained as,
\begin{equation}
\tensor{u}(\tensor{x},t)=\tensor{ \varepsilon}_M\  \tensor{x}\ \ \ 
\text{at}\ \tensor{x}\in\ \partial\vartheta.
\end{equation} 
This condition defines the linear deformation on the boundary of the RVE ($\vartheta$). 
To be specific, at each node $q$ of the surface boundary of a microstructure, we have, 
\begin{equation}
\tensor{u}_q=\tensor{\varepsilon}_M\ \tensor{x}_q\ \ \ \text{with}\ \ \ q\ =\ 1,\ \cdots,\ M,    
\end{equation}
where $\tensor{\varepsilon}_M$ is the macroscopic strain,
and $\tensor{x}_q$ is the boundary node displacement matrix of the RVE. 
In other words, one can obtain, 
\begin{equation}
    \tensor{\varepsilon}_M\colon=\left[
        \begin{matrix}{\varepsilon}_{11}&{\varepsilon}_{22}&2{\varepsilon}_{12}\\
        \end{matrix}\right]^T\ \ \ \ \ \text{and}\ \ \ \ \ 
        \tensor{u}_q\colon=\left[
            \begin{matrix}u_1&u_2\\
            \end{matrix}\right]_q^T
\end{equation}
For 2-D case, this relationship can be demonstrated as, 
\begin{equation}
\tensor{u}_q=\tensor{\mathbb{D}}_q^T\tensor{\varepsilon}_M,\ \ q=1,\cdots,\ M,
\end{equation}
where $\mathbb{D}_q$ is a matrix depends on the coordinate of nodal points in the RVE of microstructures, 
and it can be obtained as, 
\begin{equation}
\mathbb{D}_q\colon=\frac{1}{2}\left[\begin{matrix}2x_1&0\\0&2x_2\\x_1&x_2\\\end{matrix}\right]_q,
\end{equation}
where $\tensor{x}$ is defined as the displacement at the micro-level.

For homogenization of micro-level responses, 
we should partition the nodes into two groups \cite{miehe2002computational}: 
interior nodes of the RVE, $x_a\in\mathcal{V}$, and exterior nodes, $x_b\in\partial\mathcal{V}$ 
(Note: $\mathcal{V}$ is the RVE associated with microstructures). 
Therefore, we can partition the internal force vector $f\left(\tensor{u}\right)$ and 
associated tangent $K\left(\tensor{u}\right)$ of the discretized microstructure as, 
\begin{equation}
\tensor{f}=\ \left[\begin{matrix}\tensor{f}_a\\\tensor{f}_b\\\end{matrix}\right],\ \ \ \ \ \ \ 
K=\ \left[\begin{matrix}\tensor{K}_{aa}&K_{ab}\\K_{ba}&K_{bb}\\\end{matrix}\right].
\end{equation}

The homogenized tangent stiffness matrix can be obtained as, 
\begin{equation}
{\widetilde{\tensor{K}}}_{bb}=\tensor{K}_{bb}-\ \tensor{K}_{ba}\tensor{K}_{aa}^{-1}\tensor{K}_{ab} \ \ \ \text{with} \ \ \
\bar{\mathbb{C}}=\frac{1}{\left|\mathcal{V}\right|}\ \mathbb{D}{\widetilde{\tensor{K}}}_{bb}\mathbb{D}^T,
\end{equation}
where $\mathbb{C}$ is the homogenized tangent stiffness.
Finally, for the homogenization of stress, we have,
\begin{equation}
\bar{\tensor{\sigma}}=\ \frac{1}{\left|\mathcal{V}\right|}\mathbb{D}^T\tensor{f}_a,
\end{equation}
where $\left|\mathcal{V}\right|$ is the volume of the RVE of a microstructure. 

\section{Deep neural network}
The innovative idea of the ability of machines to think differently from human was developed by \citet{turing1950computing}.
A test so-called the Turing test was developed, 
in which an evaluator that differentiated between the texts was generated by the machine and human. 
During a conference in 1998, \citet{mccarthy1998artificial} practiced the term Artificial Intelligence (AI) 
as a branch of knowledge to emphasize that machines can think like a human. 
One of the subbranches of the AI is Machine Learning (ML), 
which defines a method for improving algorithms 
via experiencing new information from the database. 
As a subbranch of machine learning, considering the basic diagram of neuron \citet{mcculloch1943logical}, 
Neural networks were developed to predict simple behaviors. 
Later, Back-Propagation algorithm \cite{rumelhart1985learning} opened 
a new avenue in computer science for developing layers of neural network for several applications.
For instance, recurrent neural networks \cite{hochreiter1997long}
and convolutional neural networks \cite{lecun1999object} 
were developed to predict more complex sequence and image-based behaviors, respectively. 
By handling a large volume of data, a new subbranch, so-called Deep Learning, 
which defines a combination of several layers of neural network (more than 3) 
that is capable of extracting features and learning complex behavior, was developed. 

The overall objective of this study is to learn the heterogeneous and anisotropic 
path-dependent mechanical behaviors of materials. 
We adopt the sequence type of data that best fits Recurrent Neural Networks (RNN). 
The RNN is a type of neural networks for learning a sequential data 
and is highly applicable in the natural language process (NLP) and Speech Recognition.
For instance, the best application of the recurrent neural network is auto-completion 
technologies of Google search engine or speech recognition of Apple’s Siri. 

In this study, we implement a single Long-short term memory unit (LSTM) \cite{hochreiter1997long} 
that consists of a number of memory cells and gates for keeping (most important) 
and forgetting (less required) parts of the information in the sequence 
by minimizing the loss between target and prediction. 
It is worth noting that, despite the powerful capability of LSTM in the learning sequence of data, 
the conventional LSTM was reported not to conserve the  mass, 
which is crucial for learning path-dependent behavior \cite{hoedt2021mc}. 

\begin{figure}[!htb]
\centering
\includegraphics[trim=0cm 1cm 0cm 0cm, clip=true, width=0.8\textwidth]{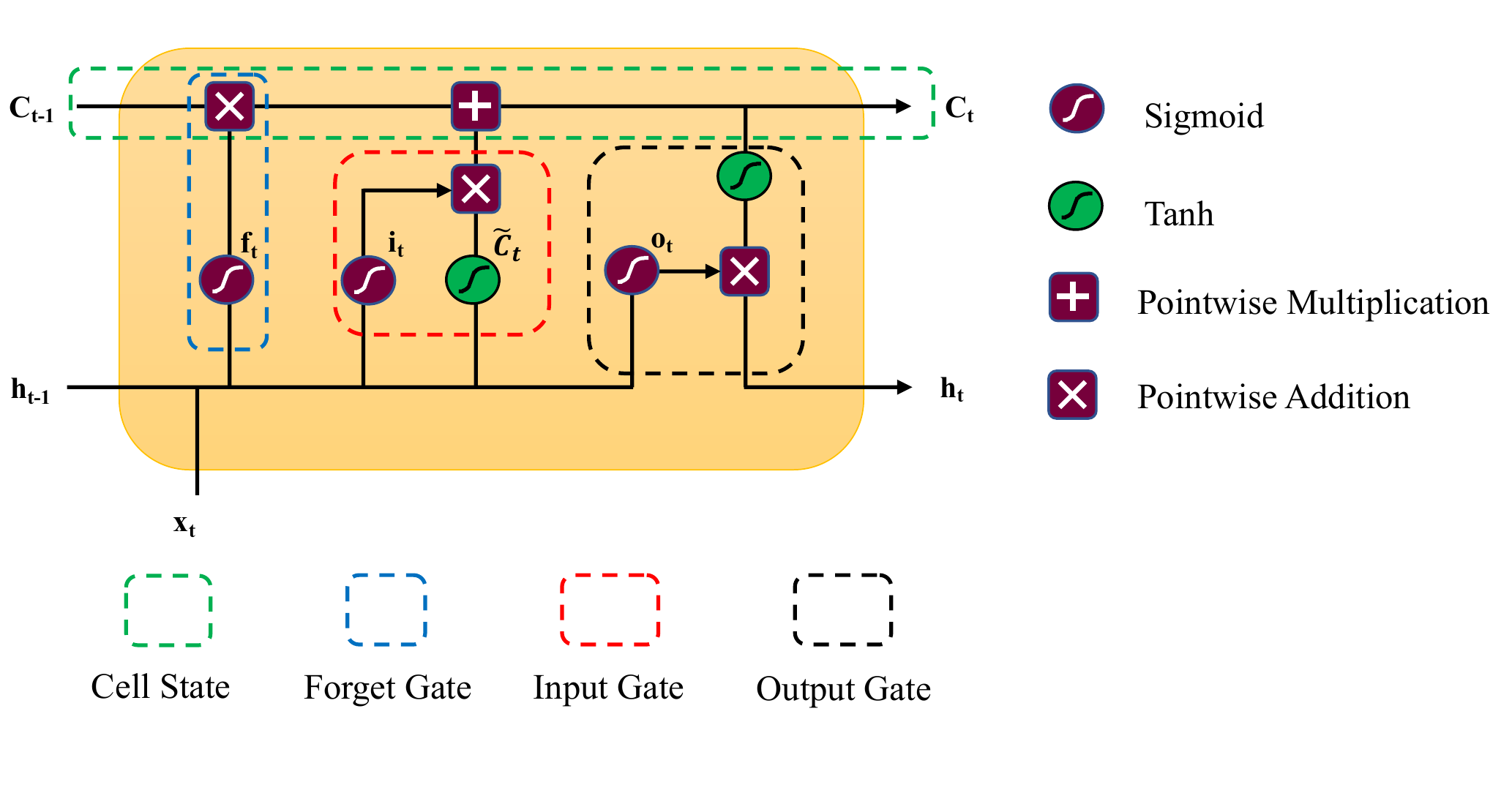}
\caption{Architecture of Long-Short Term Memory Unit (LSTM) that includes four parts: cell state, forget gate, input gate, and output gate. The LSTM decides which part of information should be preserved during the sequence to reach the minimum distance between target data and output.} 
\label{Fig: LSTM_section2}
\end{figure}

To begin with, the architecture of conventional LSTM cells are described for completeness \cite{hochreiter1997long}. 
In Figure \ref{Fig: LSTM_section2}, the Long-Short Term Unit cell consists of four components, 
input gate, forgot gate, cell state, and output gate. 
The forget cell decides which part of past information and current input should be valuable. 
The forget gate receives the current input ($x_t$) and past hidden state ($h_t$),
which are passed through the sigmoid function. 
The information mapped near one is valuable and preserved. 
The forget gate can be obtained as, 
\begin{equation}
f_t=\sigma(W_f.[h_{t-1}, x_t]+b_f),
\end{equation}
where $t$, $f_t$, $x_t$, $h_{t-1}$, $W_f$, and $b_f$ are the time-step, forget gate at $t$, 
input, previous hidden state, weight matrix between forget and input gate, and connection 
bias at $t$, respectively. 
Finally, the value of $f_t$ is considered in cell state with point-wise multiplication. 

The input gate executes two operations: (1) the current input $x_t$ and the previous hidden 
state $h_{t-1}$ mapped between 0 (not valuable) and 1 (valuable) using sigmoid function;
(2) the similar current input and previous hidden states pass through tanh function to 
regulate the network by creating a vector ${\widetilde{C}}_t\left(t\right)$ between -1 and 1. 
These two outputs are combined with point-wise multiplication and added to the cell state. 
The two operations of the input gate can be given as, 
\begin{equation}
\begin{gathered}
i_t=\sigma(W_i.[h_{t-1}, x_t])+b_i \ \ \ \text{and} \ \ \
{\widetilde{C}}_t=\tanh(W_c.[h_{t-1},x_t]+b_c),
\end{gathered}
\end{equation}
where $t$, $i_t$, $W_i$, $b_i$ are the time-step, input gate at t and weight matrix of sigmoid 
operation between input and output gate, respectively. 
${\widetilde{C}}_t$, $W_c$, and $b_c$ are value generated by tanh, weight matrix of tanh, and bias vector at $t$. 

The cell state performs as the memory of the LSTM unit, and it includes two operations: 
first, the value of previous cell state $C_{t-1}$ is multiplied to the forget vector to decide
what part of past cell state information should be dropped;
second, the network update the cell state with adding the point-wise multiplication of input vector ($i_t$) 
with vector (${\widetilde{C}}_t$). 
The cell state can be given as, 
\begin{equation}
C_t=f_t\ast C_{t-1}+i_t\ast{\widetilde{C}}_t.
\end{equation}
Finally, the output gate determines the value for the next hidden state by point-wise 
multiplication of the output gate ($o_t$) and the value of the current cell state passed 
through tanh function. The forget cell functions can be given as, 
\begin{equation}
\begin{gathered}
o_t=\sigma(W_o.[h_{t-1},x_t])+b_o \ \ \ \text{and} \ \ \
h_t=o_t\ast \tanh(C_t),
\end{gathered}
\end{equation}
where $o_t$ is output gate at t and $h_t$ is the hidden state of current step.

\section{Design of experiment}
\label{sec:DOE}
This section explains the design of virtual experiments to generate a database for 
learning heterogeneous and anisotropic responses of a solid.
To present the robustness of Deep Learning 
in capturing the path-dependent behaviors with heterogeneous and anisotropic features,
this section is divided into three subsections:
(1)  Generation of Loading Path: Definition of generated loading paths 
to be used for identifying target responses of materials; 
(2) Generation of heterogeneous Database: 
Description of considered heterogeneity with randomness for constitutive laws; 
(3) Generation of anisotropic Database: Description of horizontal layers 
for explicit representation of anisotropic microstructures (transversely isotropic) 
and collection of homogenized microstructural responses undergoing diverse loading paths. 
   
\subsection{Generation of loading path}\label{subsec:generation_loading_path}
For capturing the path-dependent behavior via neural networks, 
a supervised learning method is implemented,
where a set of input, strain path and features, is mapped to a set of target data, stress path. 
Therefore, generation of a database that includes a set of strain and stress loading paths is an essential part for the learning process. 
Besides, the database should be general enough to train the networks by material responses
instead of allowing the networks to over fit the biased behaviors. 
For this purpose, random generation of loading path is considered in this study \cite{mozaffar2019deep}. 

\begin{figure}[!htb]
    \centering
    \subfigure[]{\includegraphics[trim=1cm 13cm 6cm 6cm, clip=true,width=0.45\textwidth]{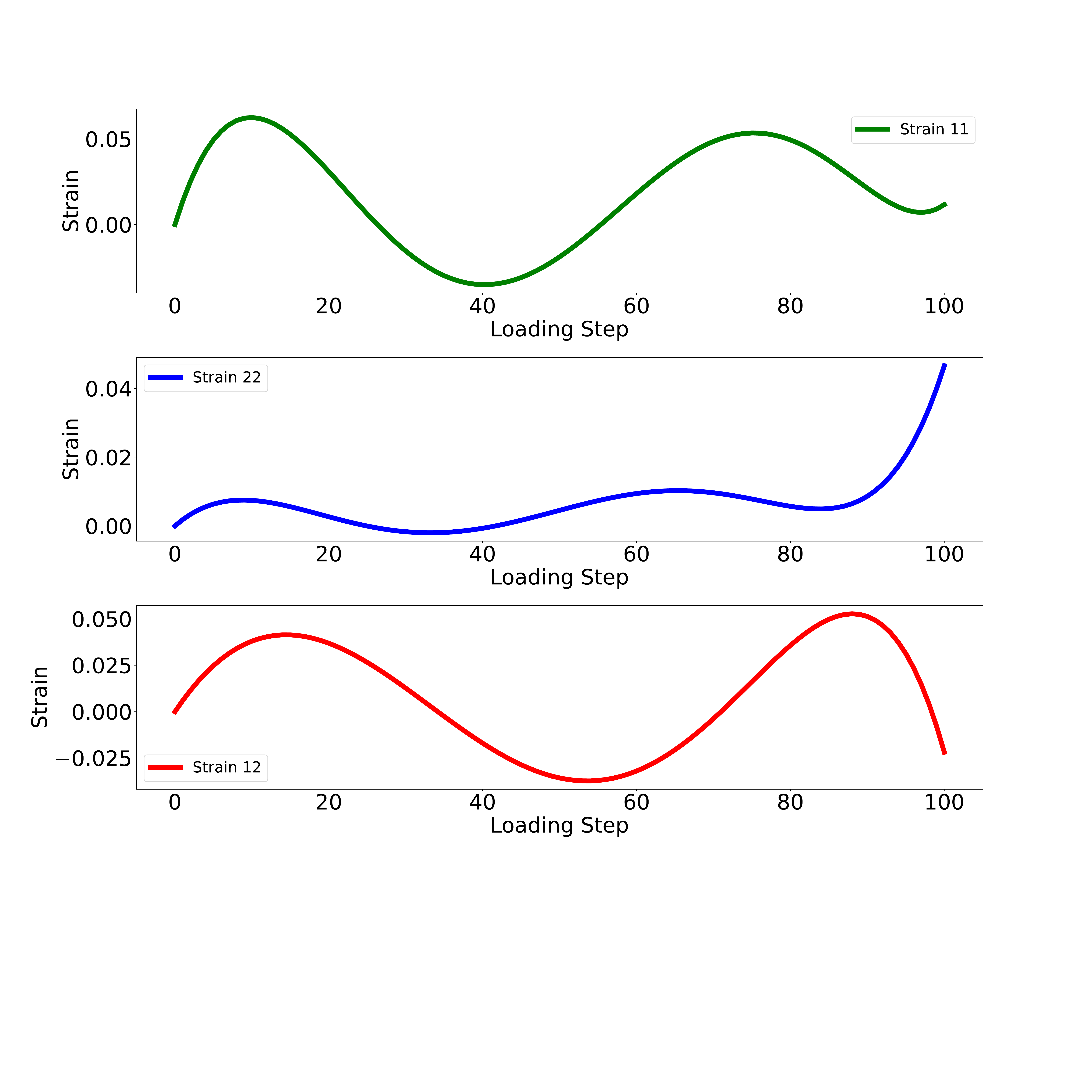}\label{fig-randstrain1a}} 
    \subfigure[]{\includegraphics[trim=1cm 13cm 6cm 6cm, clip=true,width=0.45\textwidth]{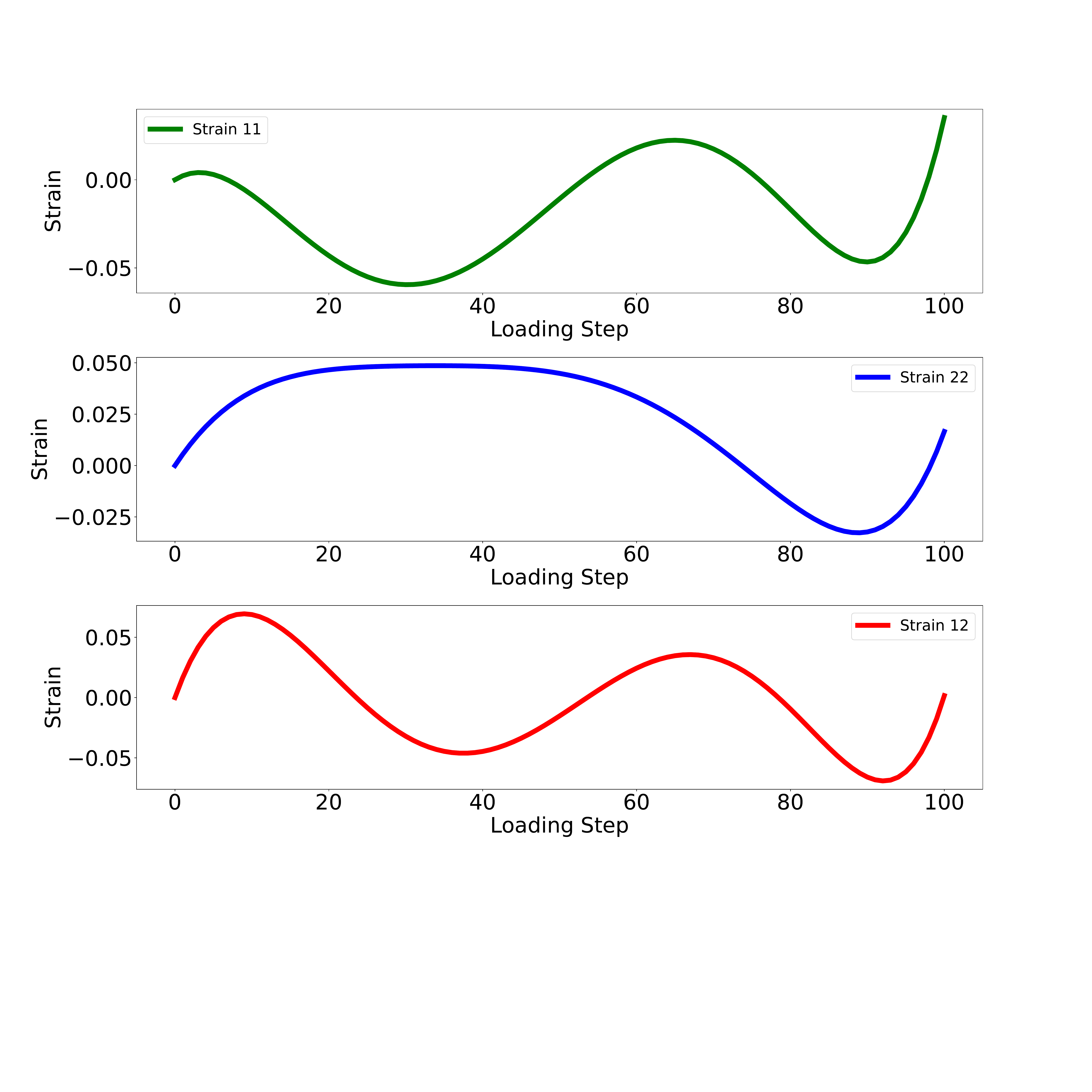}\label{fig-randstrain1b}}        
    \caption{Two samples of randomly generated loading paths in Section \ref{subsec:generation_loading_path}}
    \label{Fig: RadnomLoadingPath_section2}
\end{figure}

The random generation method for mechanical loading path may change the static nature of the problem.
In other words, non-physical oscillatory patterns can be imposed into the loading path when the random generation is used directly. 
To remedy this potential issue, the following method is designed. 
First, the loading path is considered with 100 loading steps, which are constituted by random strain values within a target strain range. 
Among the loading steps, strain values of 0, 20, 40, 60, 80, and 100 steps are only selected. 
Then a six-degree polynomial that is fitted into those  six strain values is introduced
to fill the rest of strain values for each loading step. 
Finally, the initial loading step is set to zero for simplicity. 
Figure \ref{Fig: RadnomLoadingPath_section2} depicts two examples of randomly generated loading paths 
with the strain range of (-0.05-0.05).   

\subsection{Generation of database with material heterogeneity}
Introducing heterogeneity into a constitutive law is described
to investigate the applicability of a single Long-Short Term Memory Unit associated with path-dependent responses of solids.
Mechanical responses against randomly generated loading paths are considered
using the J2 plasticity with isotropic hardening.
In this problem, a single LSTM network architecture is demonstrated to learn path-dependent constitutive behaviors 
associated with material heterogeneity. 
The material heterogeneity is defined by adopting various material properties
including elastic properties, yield stress, and hardening parameter. 

Table \ref{tab: heterogenity_problem1} presents the ranges of each material parameter to account for heterogeneity
associated with the J2 plasticity model with isotropic hardening. 
14,000 sets of material properties are considered to construct a database,
where four material properties are randomly selected. 
Then, random loading paths (Section \ref{subsec:generation_loading_path})
are applied to collect the components of stress tensors for each of 14,000 loading paths. 
Thus, the following sets of data are collected for the database, that are,
(1) loading paths with selected material properties;
(2) stress responses under each loading path. 
For more information about learning heterogeneity, please refer to Section \ref{sec : problem2}.

\begin{table}[!htb]
	\caption{Ranges of material properties for heterogeneity in J2 plasticity with isotropic hardening}
	\label{tab: heterogenity_problem1}
	\begin{center}
		\begin{tabular}{cc}
			\hline
			Material Properties & Value (GPa) \\
			\hline
			$\lambda$ & 200-400 \\
			$\mu$ & 20-40 \\
			$H$ & 20-40 \\
			$\kappa$ & 0.2-0.4\\
			\hline
		\end{tabular}
	\end{center}
\end{table}

\subsection{Generation of  database with anisotropy}
Anisotropy is often observed from materials with layers, such as composites, soils, and rocks. 
In this study, a simplified layered microstructure representing transversely isotropic feature is considered.
The database is constructed by collecting homogenized responses 
of such anisotropic microstructures the via FE$^2$ framework. 
For simplicity, two-dimensional microstructures with two different embedded materials,
that are a host matrix with inclusions, is designed to represent transverse isotropy.
The host matrix is considered to be a linear elastic material (Material 1)
while the constitutive law of inclusions is assumed to be elasto-plastic with the J2 plasticity (Material 2)
Material properties of each of Material 1 and 2 are presented 
in Table \ref{tab: material anisotropy_section3}.

\begin{table}[!htb]
	\caption{Material properties of the layered microstructure (GPa): Material 1 for the host matrix and Material 2 for inclusions}
	\label{tab: material anisotropy_section3}
	\begin{center}
		\begin{tabular}{ccccc}
			\hline
			Material Type & $\lambda$ & $\mu$ & $H$ & $\kappa$  \\
			\hline
			Material 1 & 51.0836 & 26.3158 & - & -\\
			Material 2 & 1.8244 & 0.9398 &  0.2 & 0.039 \\
			\hline
		\end{tabular}
	\end{center}
\end{table}

\begin{figure}[!htb]
	\centering
	\subfigure[]{\includegraphics[width=0.3\textwidth]{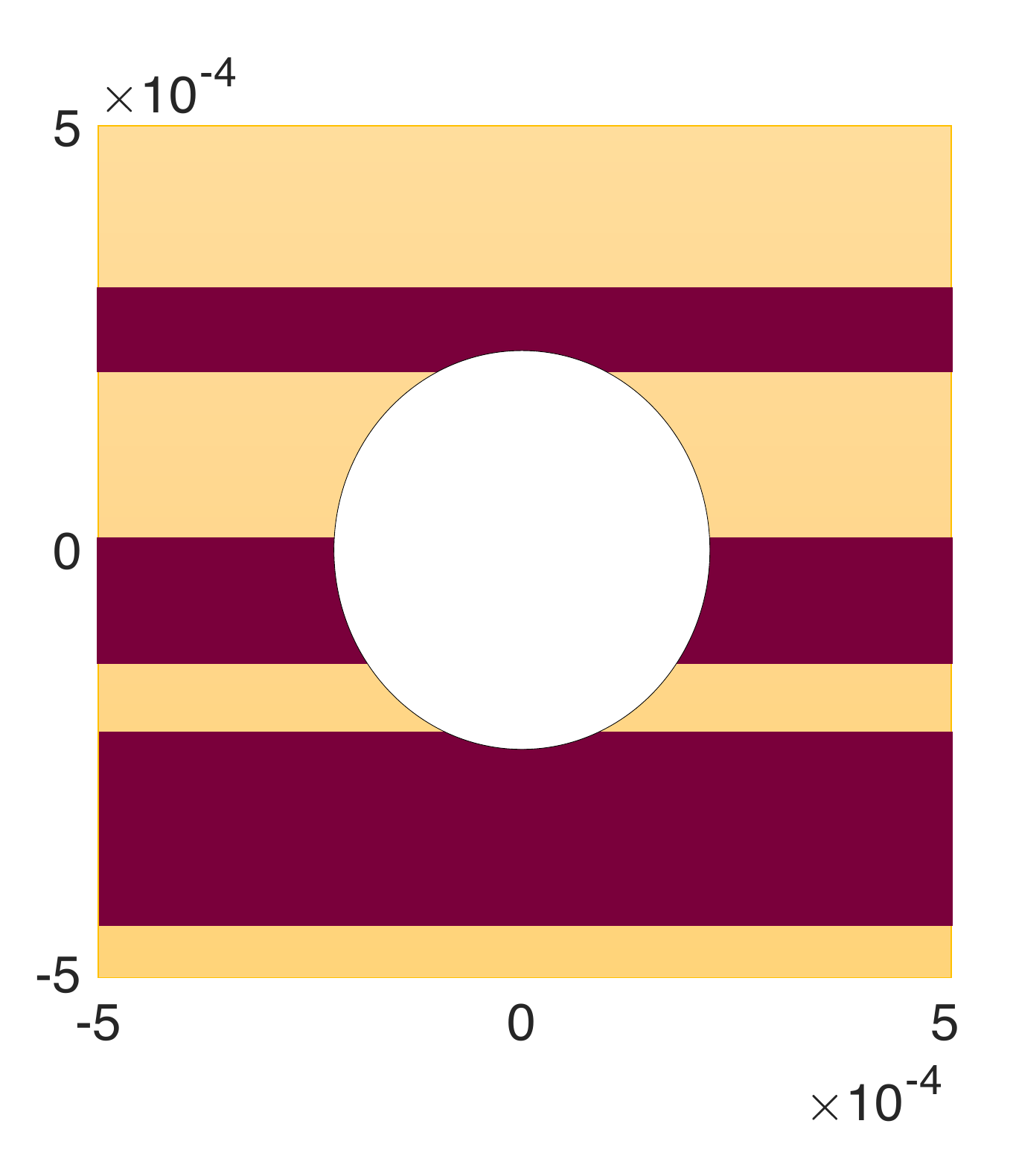}\label{fig-anisomicro1}} 
	\subfigure[]{\includegraphics[width=0.3\textwidth]{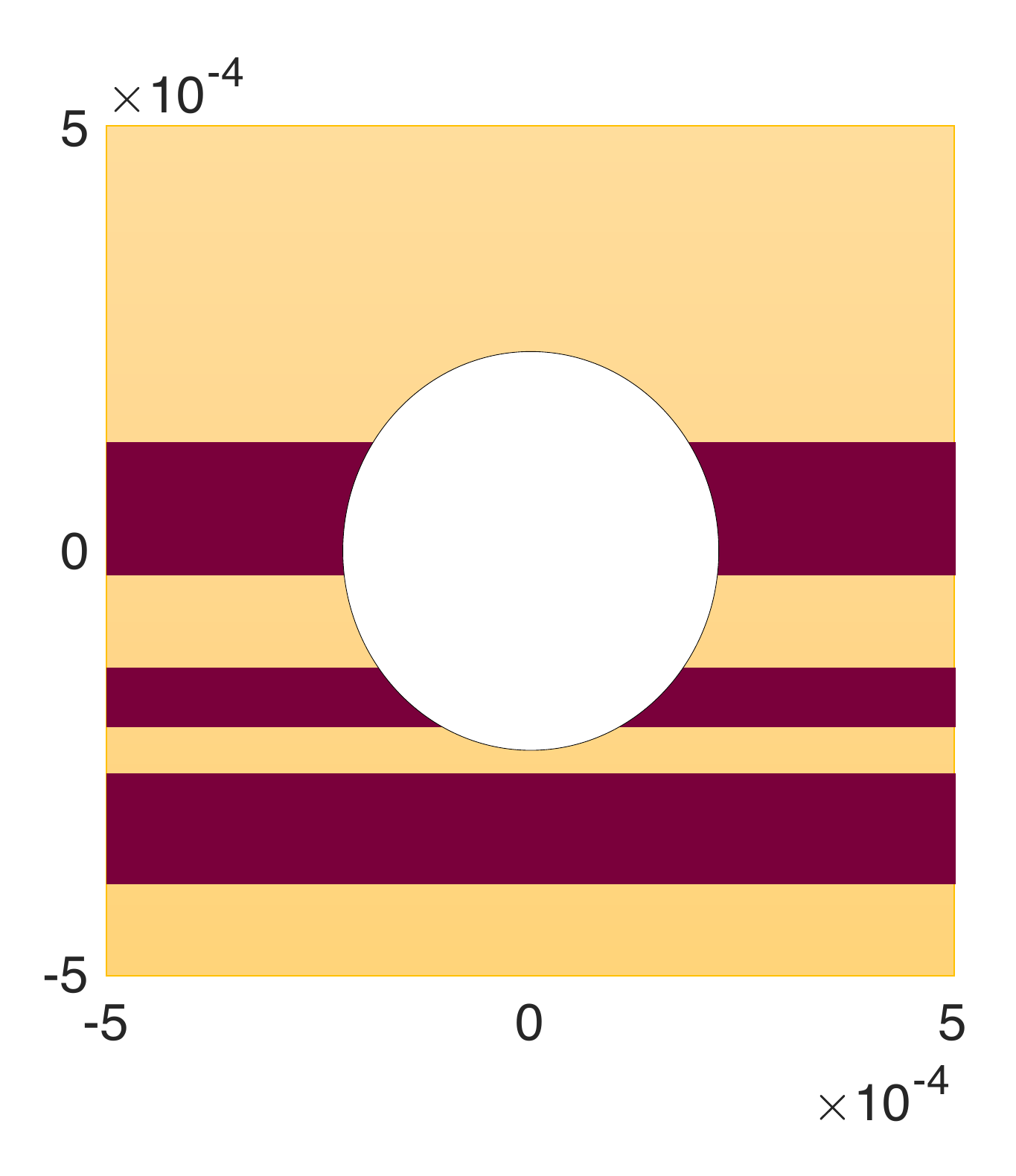}\label{fig-anisomicro2}} 
	\subfigure[]{\includegraphics[width=0.3\textwidth]{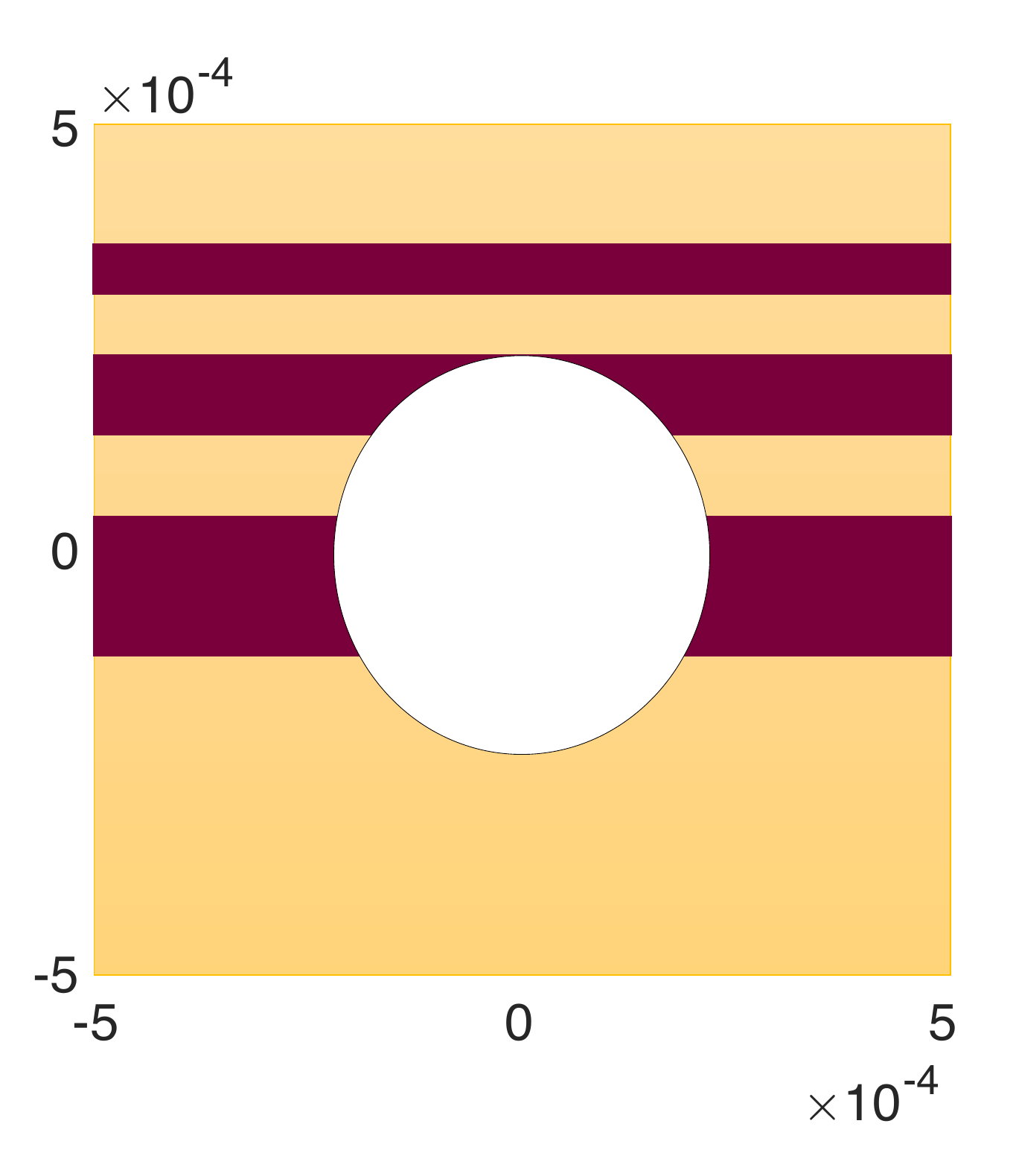}\label{fig-anisomicro3}} 
	\caption{Schematic representation of heterogeneous anisotropic microstructures (Material 1 for the host matrix and Material 2 for the inclusions (Table \ref{tab: material anisotropy_section3})): (a) Micro 1, (b) Micro 2, (c) Micro 3, respectively}
	\label{Fig: anisomicro_section3}
\end{figure}

A unit cell of the representative volume element (RVE)
has a square domain with the length of 1 $\mu m$ including a void with the volume fraction of 15 percent with respect to the unit cell. 
Each cell has seven horizontal layers with alternate material properties of Materials 1 and 2.
We limit our analysis with three embedded layers that are considered as inclusions with Material 2. 
In this problem, geometrical heterogeneity of transversely isotropic microstructures
is considered by randomly generating horizontal lines in a unit cell,
which differentiates the thickness and location of each layer. 
It is worth noting that the minimum thickness of each layer is set to 0.05 $\mu m$.
Figures \ref{Fig: anisomicro_section3} depicts three samples of randomly configured microstructures.

After generating transversely isotropic microstructures, 
three descriptors are defined to identify the heterogeneity of each anisotropic microstructure,
which are (1)  the relative thickness of three inclusion layers, 
(2) the volume fraction of Material 2, and
(3) the location of three inclusion layers from the bottom of each unit cell. 
In other words, these descriptors are extracted to connect the mechanical responses 
with anisotropic heterogeneity of each microstructure. 
For example, Table \ref{tab: descriptor_aniso_section3} presents the descriptors of
those sample microstructures in Figure \ref{Fig: anisomicro_section3}.
Again, random loading paths explained in Section \ref{subsec:generation_loading_path} are applied 
to randomly configured anisotropic microstructures,
in which homogenized stress responses are collected 
to construct a database for training deep neural networks.

\begin{table}[!htb]
\caption{Descriptors of three anisotropic microstructures in Figure \ref{Fig: anisomicro_section3}}
\label{tab: descriptor_aniso_section3}
 \begin{center}
 \begin{tabular}{cccc}
   \hline
   Descriptor & Micro  1 & Micro  2 & Micro  3  \\
   \hline
  Relative Fraction of Layers & 0.48  & 0.347 & 0.314 \\
  Relative Thickness & [0.229, 0.151, 0.099] & [0.127, 0.0679, 0.152] & [0.162, 0.093, 0.057]  \\
  Relative Location & [0.054, 0.364, 0.709] & [0.105	, 0.294, 0.469] & [0.373, 0.636, 0.803]  \\
  \hline
\end{tabular}
\end{center}
\end{table}

\section{Verification and results}
\label{sec:problems}
Three parts are designed to investigate the capability of a single LSTM network architecture for capturing path-dependent behavior
using the database discussed in Section \ref{sec:DOE}. 
First, a benchmark simulations is selected first 
to verify the FE$^2$ framework proposed in Section \ref{sec : FEM2}. 
In the second part, the basic LSTM is demonstrated to learn
the heterogeneous path-dependent behaviors, that is followed by the J2 plasticity model. 
The third part explores applicability of the conventional LSTM
for learning homogenized microstructural responses, 
which consider path-dependent anisotropic behavior with geometrical heterogeneity of microstructures. 

\subsection{Part I: Verification of FE$^2$ Framework}
The double scale finite element method (FE$^2$) is a concurrent multiscale framework, where 
both macro and micro scales are discretized followed by the finite element method. 
The FE$^2$ approach is a powerful tool for capturing heterogeneity and anisotropy of microstructures,
which adopts the computational homogenization for overall behaviors of microstructures. 
The governing equations for FE$^2$ framework are presented in Section \ref{sec : FEM2}. 
To verify the proposed FE$^2$ framework, a unit cell with the square domain 
for the representative volume element (RVE) at the micro-scale is considered followed by \citet{peric2011micro}.
The isotropic hardening J2 plasticity model is assigned to for the cell matrix, 
that includes a void with the volume fraction of 15 percent with respect to the unit cell. 
The material properties for the matrix are: Young’s modulus $E=70$ GPa, 
Possion’s ratio $\nu = 0.2$, initial yield stress $\sigma_0 = 0.243$ GPa, 
and hardening modulus $H = 0.2$ GPa. 

Computational homogenization is modeled under the plane stress assumption within the small strain regime.
The macro strain is applied over the RVE, and the initial boundary value problem is solved via the proposed
boundary condition \cite{peric2011micro}. The macro strain is defined as: 
\begin{equation}
[{\bar{\varepsilon}}_{11},{\bar{\varepsilon}}_{22},{\bar{\varepsilon}}_{12}]=[0.001,0.001,0.0034]
\end{equation}

\begin{figure}[h!]
	\centering
	\includegraphics[width=.45\textwidth]{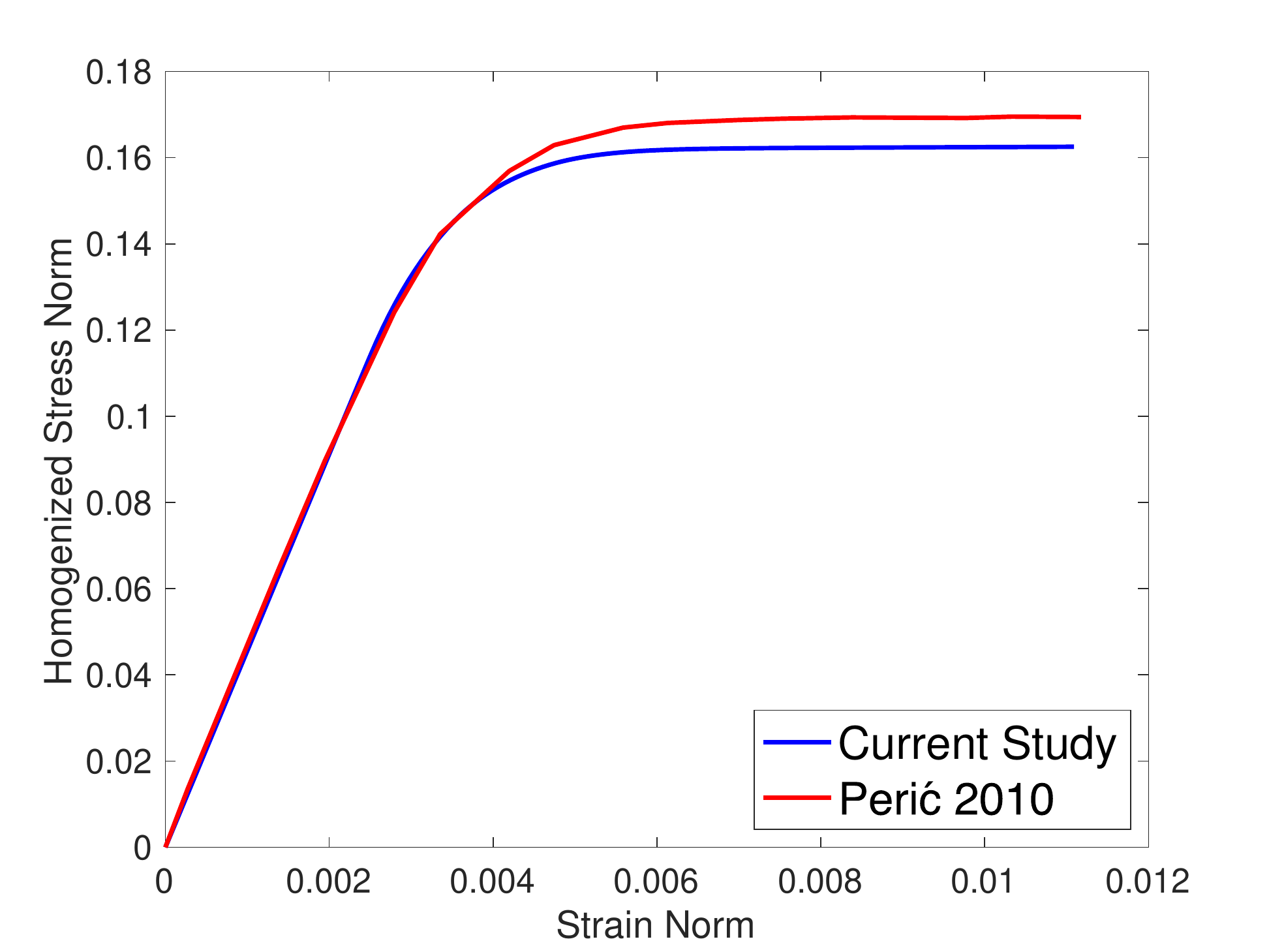}
	\caption{Benchmark test for the FE$^2$ concurrent multiscale framework followed by \citet{peric2011micro}} 
	\label{Fig: verification_problem0}
\end{figure}

\begin{figure}[h!]
    \centering
    \subfigure[]{\includegraphics[width=0.3\textwidth]{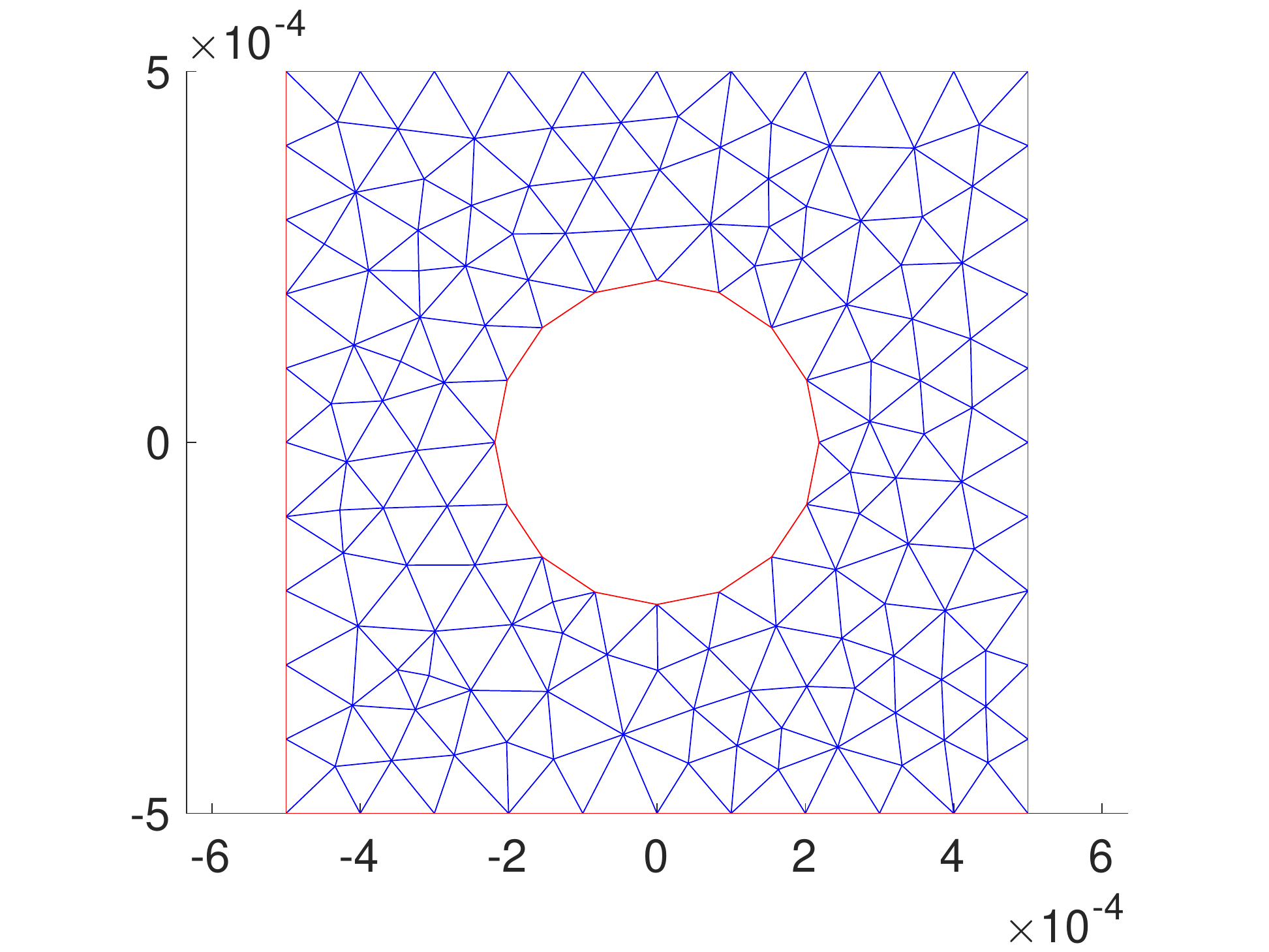}\label{fig-cellprob0a}} 
    \subfigure[]{\includegraphics[width=0.3\textwidth]{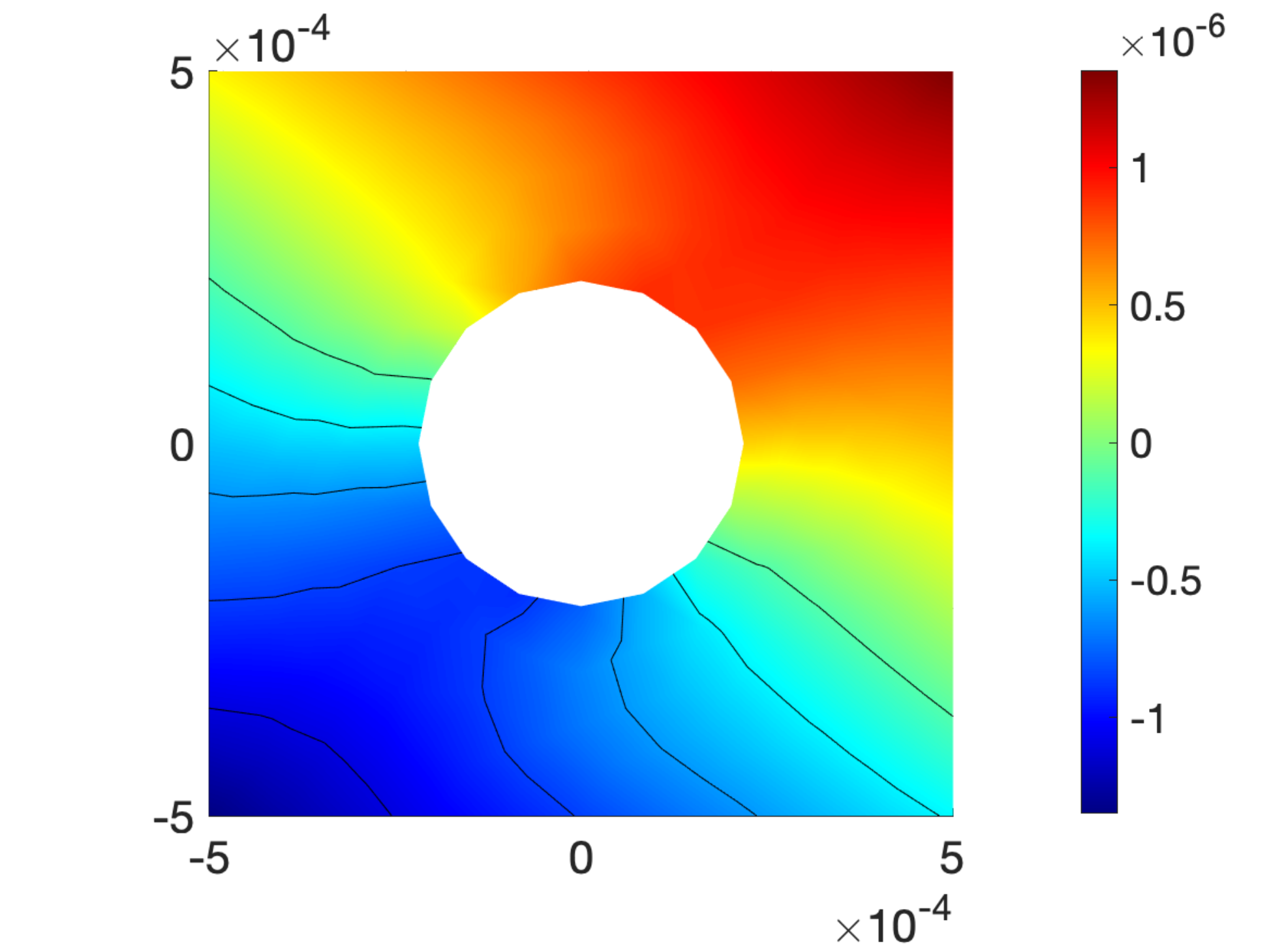}\label{fig-cellprob0b}} 
    \subfigure[]{\includegraphics[width=0.3\textwidth]{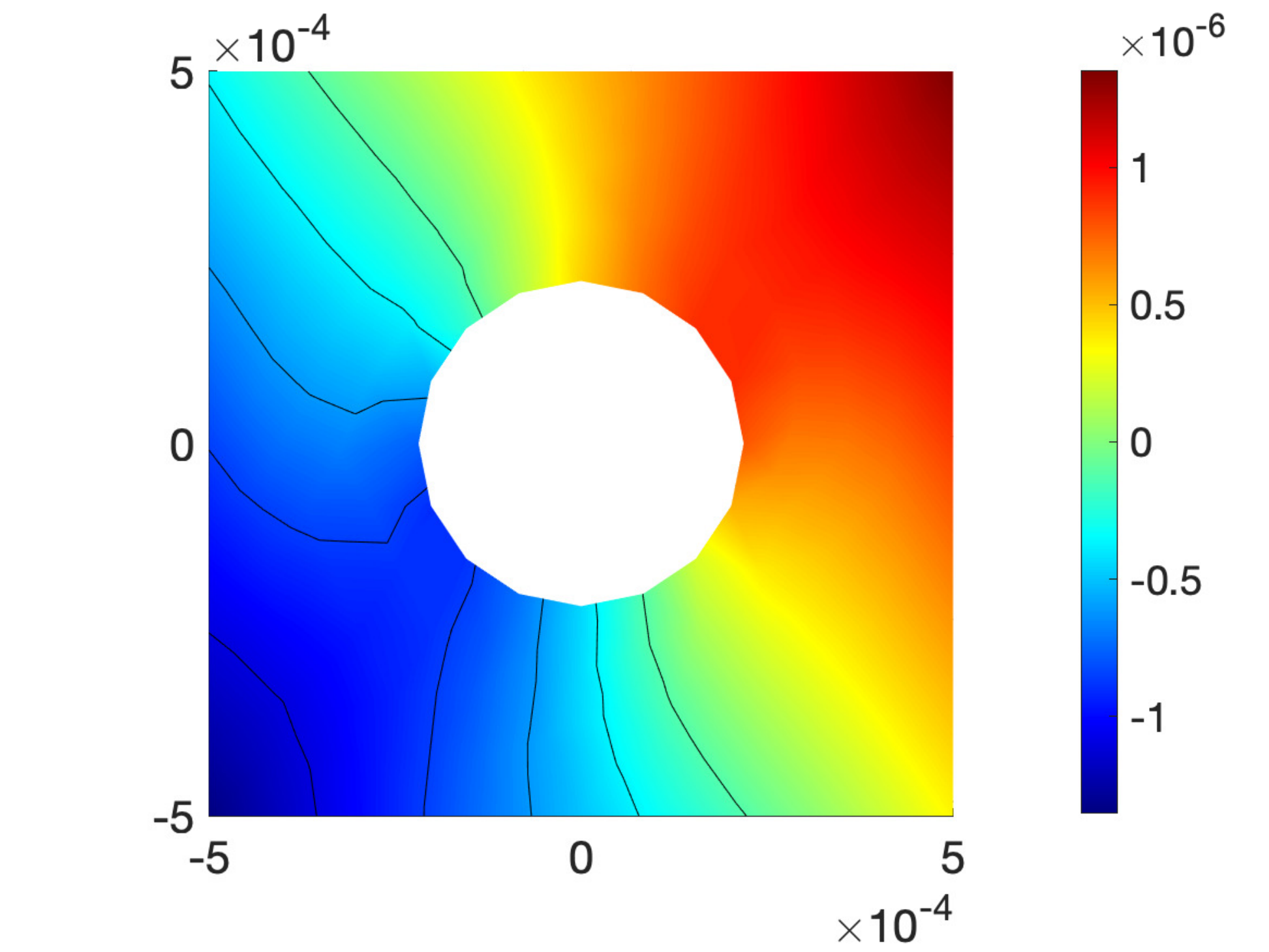}\label{fig-cellprob0c}} 
    \caption{Configuration of a microstructure for the benchmark test in Figure \ref{Fig: verification_problem0} and its results:
    	(a) Mesh, (b) X Displacement, (c) Y Displacement - the units are in m}
    \label{Fig: MicroCell_Problem0}
\end{figure}

The macro strain proportionally increases by multiplying a loading factor to the above generic stress. 
Figures \ref{Fig: verification_problem0} and  \ref{Fig: MicroCell_Problem0} depict 
the verification of FE$^2$ framework and the results of simulations, respectively. 

\subsection{Part 2: Learning heterogeneous J2 plasticity}
\label{sec : problem2}
Path-dependent behavior may originate from the permanent irregularity within the crystal structure 
(dislocation) under loading \cite{borja2013plasticity}.
The complexity of developing a mathematical framework to predict heterogeneous path-dependent behavior 
demonstrates the desire to use the Deep Learning method to predict path-dependent behaviors
\cite{mozaffar2019deep,fuchs2021dnn2}.
Therefore, the overall objective of this part is 
to investigate a procedure of learning heterogeneous path-dependent behaviors
associated with constitutive laws. 
Among several constitutive laws for modeling path-dependent behaviors,
this study focuses on the isotropic hardening J2 plasticity model \cite{borja2013plasticity}.
The heterogeneity is considered on the path-dependent behavior 
by changing material properties followed by Table \ref{tab: heterogenity_problem1}.    

The initial step for training deep neural networks is the data generation. 
A set of stress responses is obtained by simulating the constitutive model
with different ranges of material properties as in Table \ref{tab: heterogenity_problem1} 
under random loading paths. 
14,000 sets of heterogeneous material properties are considered for this problem,
which may imply samples of microstructures with overall heterogeneous material properties.
As described in Section \ref{sec:DOE}, randomly generated loading paths are applied to a sample, 
and the corresponding stress responses are collected via the return mapping algorithm \cite{borja2013plasticity}.
A set of loading paths, material properties, and stress responses 
is collected for 14,000 samples. 

\begin{figure}[!htb]
	\centering
	\includegraphics[width=1\textwidth]{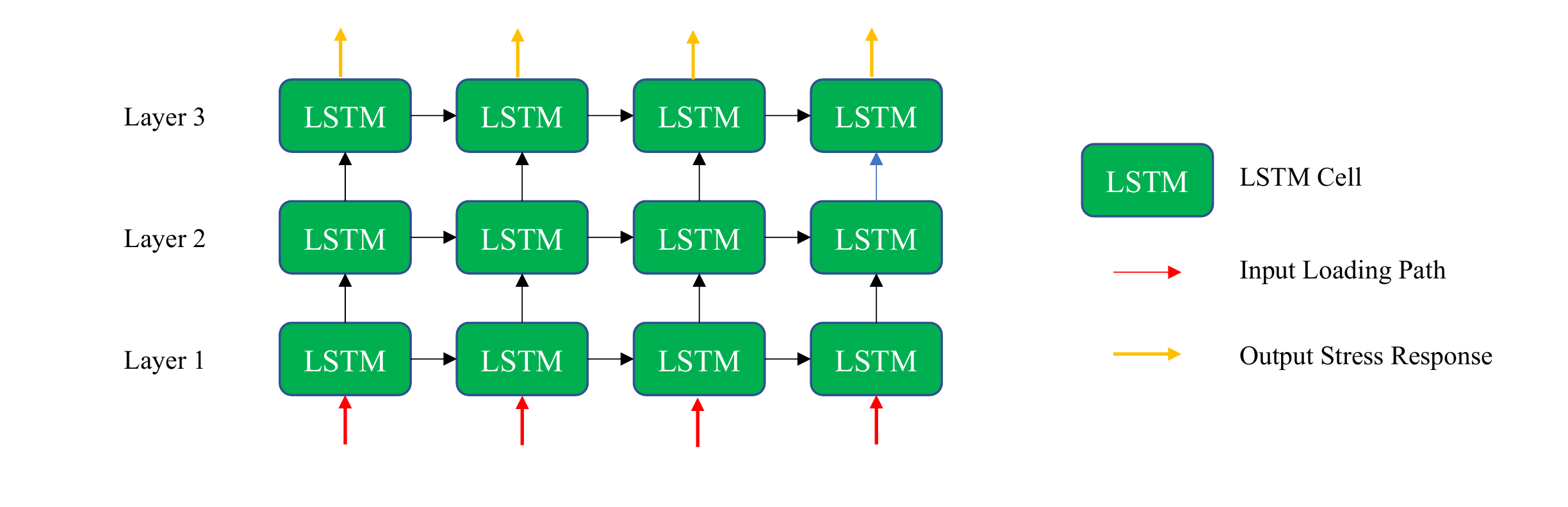}
	\caption{Architecture of Deep Recurrent Neural Networks, stacked with Long-Short Term Memory (LSTM) units} 
	\label{Fig: DeepNetArchitecture_problem1}
\end{figure}

The Architecture of the Deep Network consists of 3 stacked layers of long-short term memory units (LSTM),
 as depicts in Table \ref{tab: ArchitectureofNetwork_problem1} and Figure \ref{Fig: DeepNetArchitecture_problem1}. 

\begin{table}[!htb]
	\caption{Architecture of the Long-Short Term Memory (LSTM) Unit for Part 2 (Section \ref{sec : problem2})}
	\label{tab: ArchitectureofNetwork_problem1}
	\begin{center}
		\begin{tabular}{ccc}
			\hline
			Layer (Type) & Output Shape  & Activation Function\\
			\hline
			Input & (None, 101, 10) & None  \\
			LSTM & (None,101,250) & tanh \\
			LSTM & (None,101,250) & tanh \\
			LSTM & (None,101,250) & tanh\\
			Time Distributed & (None, 101, 3) & LeakyReLU  \\
			\hline
		\end{tabular}
	\end{center}
\end{table} 

\begin{figure}[!htb]
    \centering
    \subfigure[]{\includegraphics[width=0.3\textwidth]{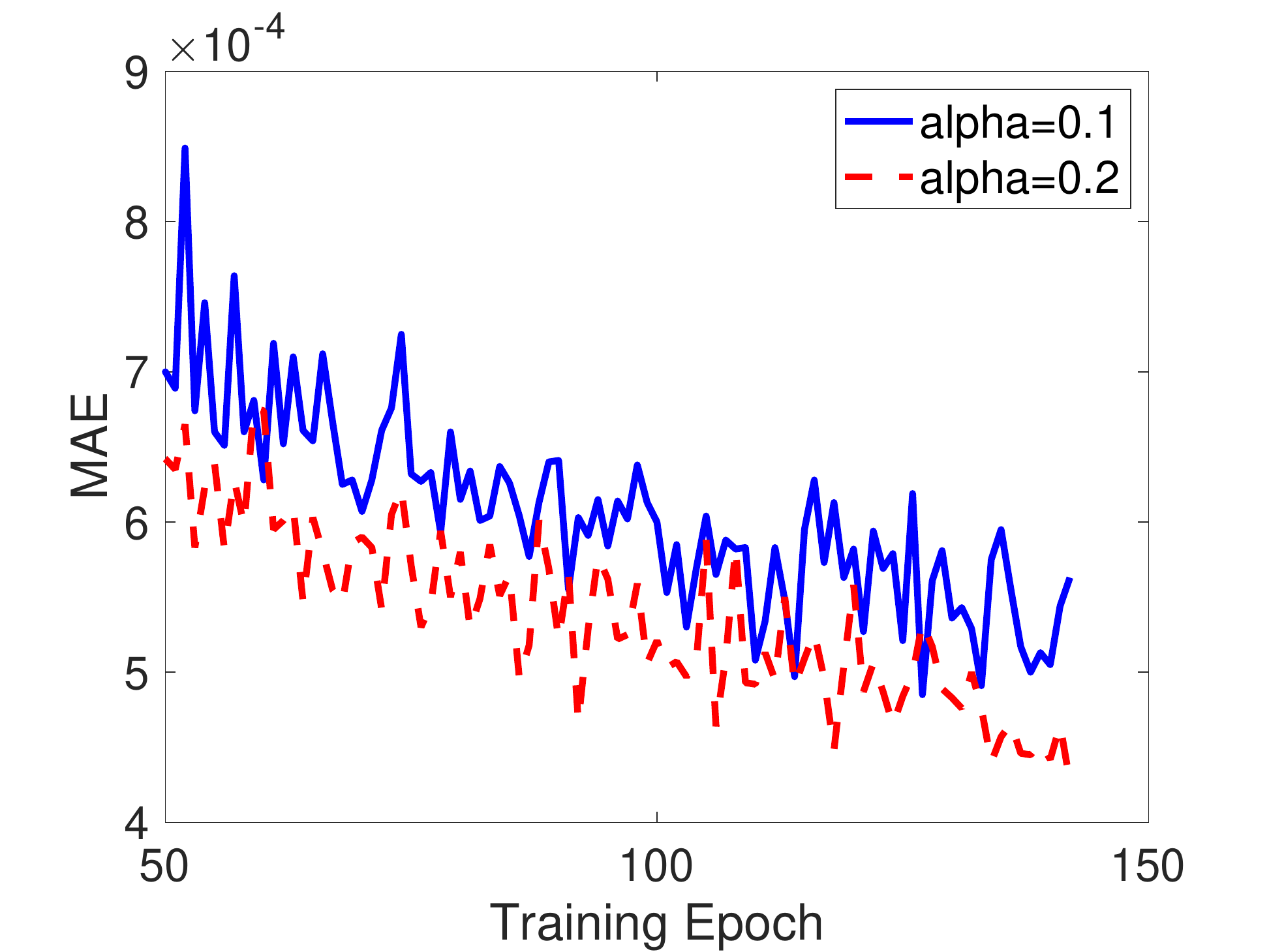}\label{fig-prob1a}} 
    \subfigure[]{\includegraphics[width=0.3\textwidth]{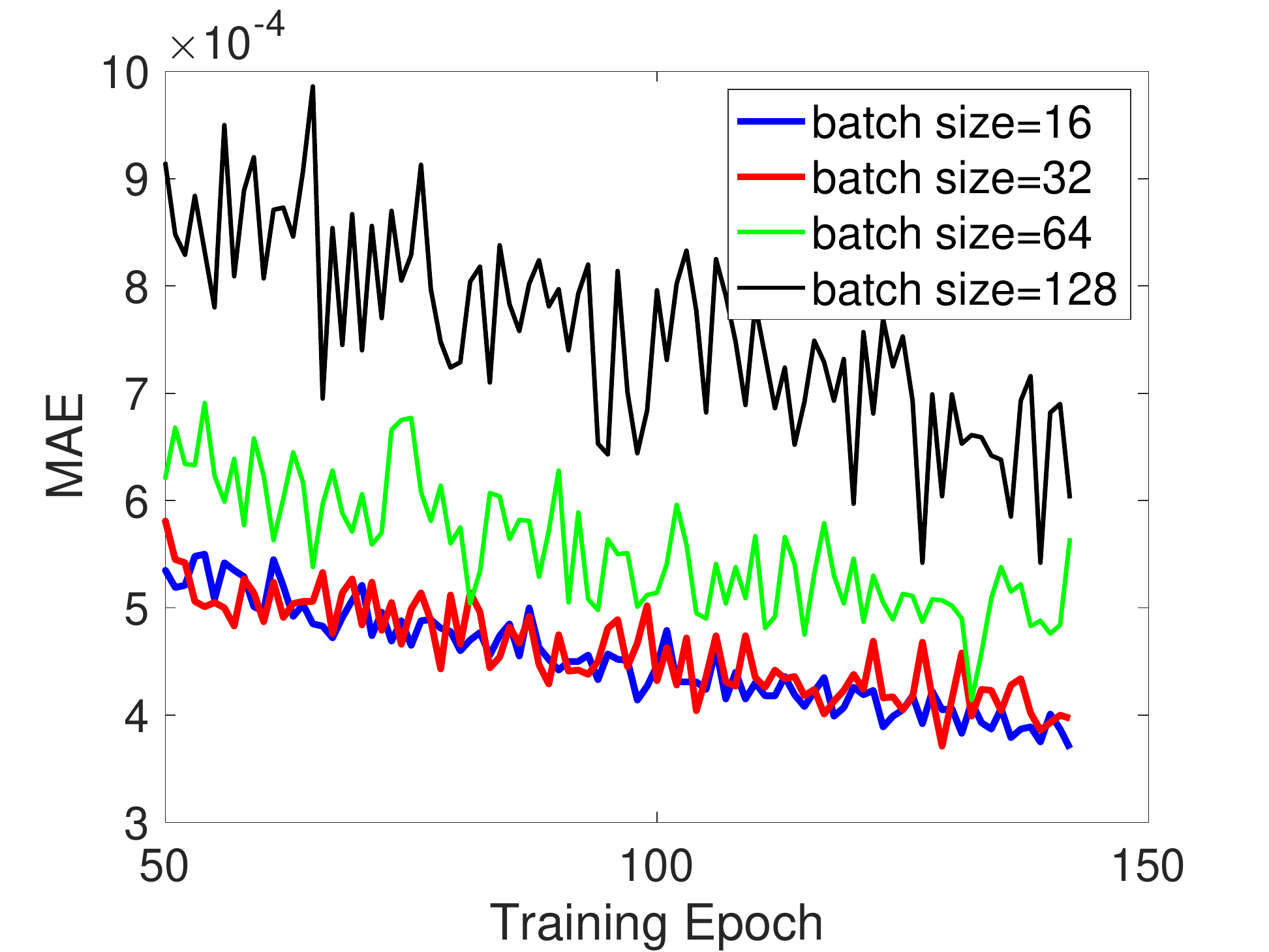}\label{fig-prob1b}} 
    \subfigure[]{\includegraphics[width=0.3\textwidth]{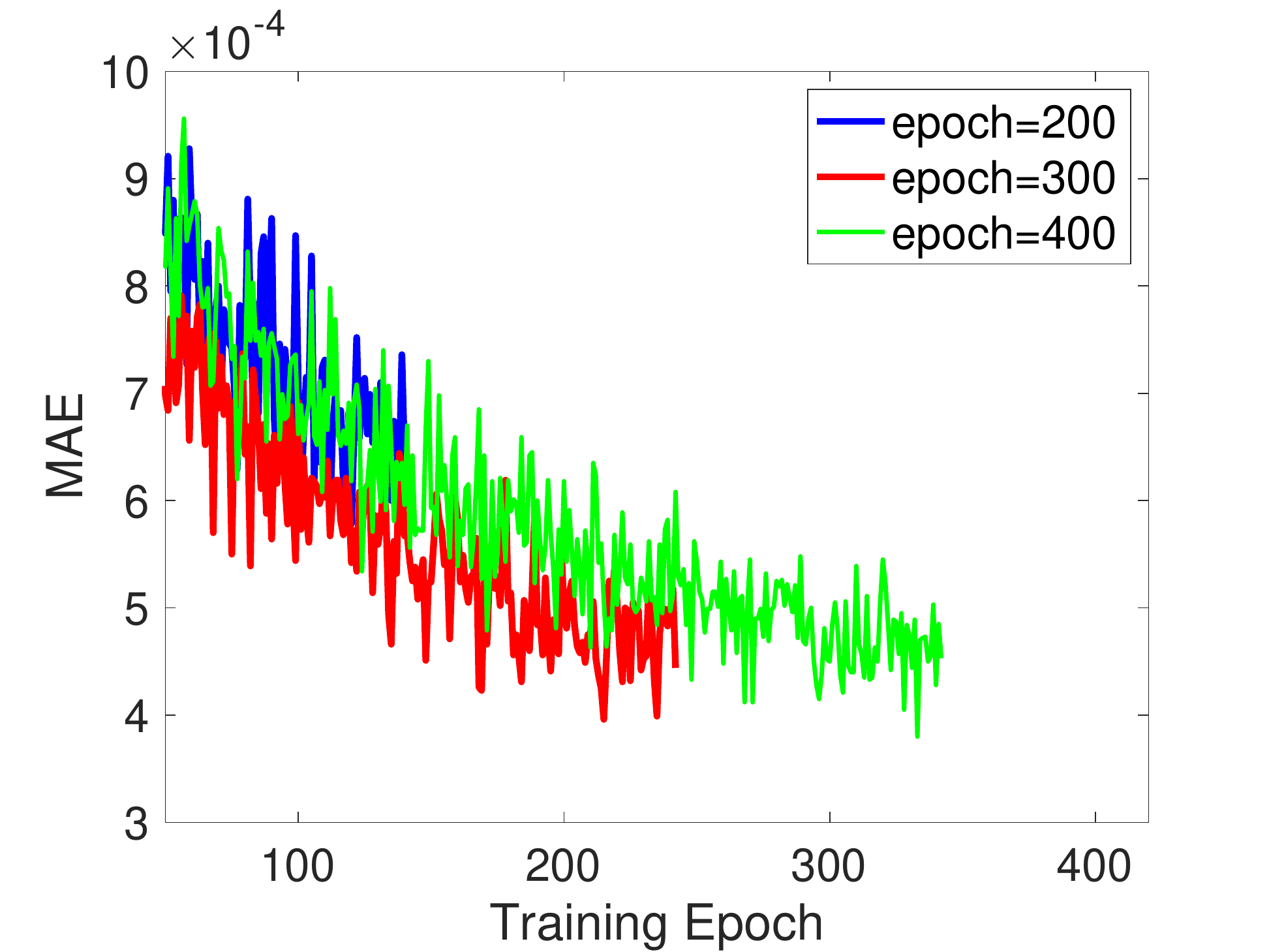}\label{fig-prob1c}} 
    \subfigure[]{\includegraphics[width=0.3\textwidth]{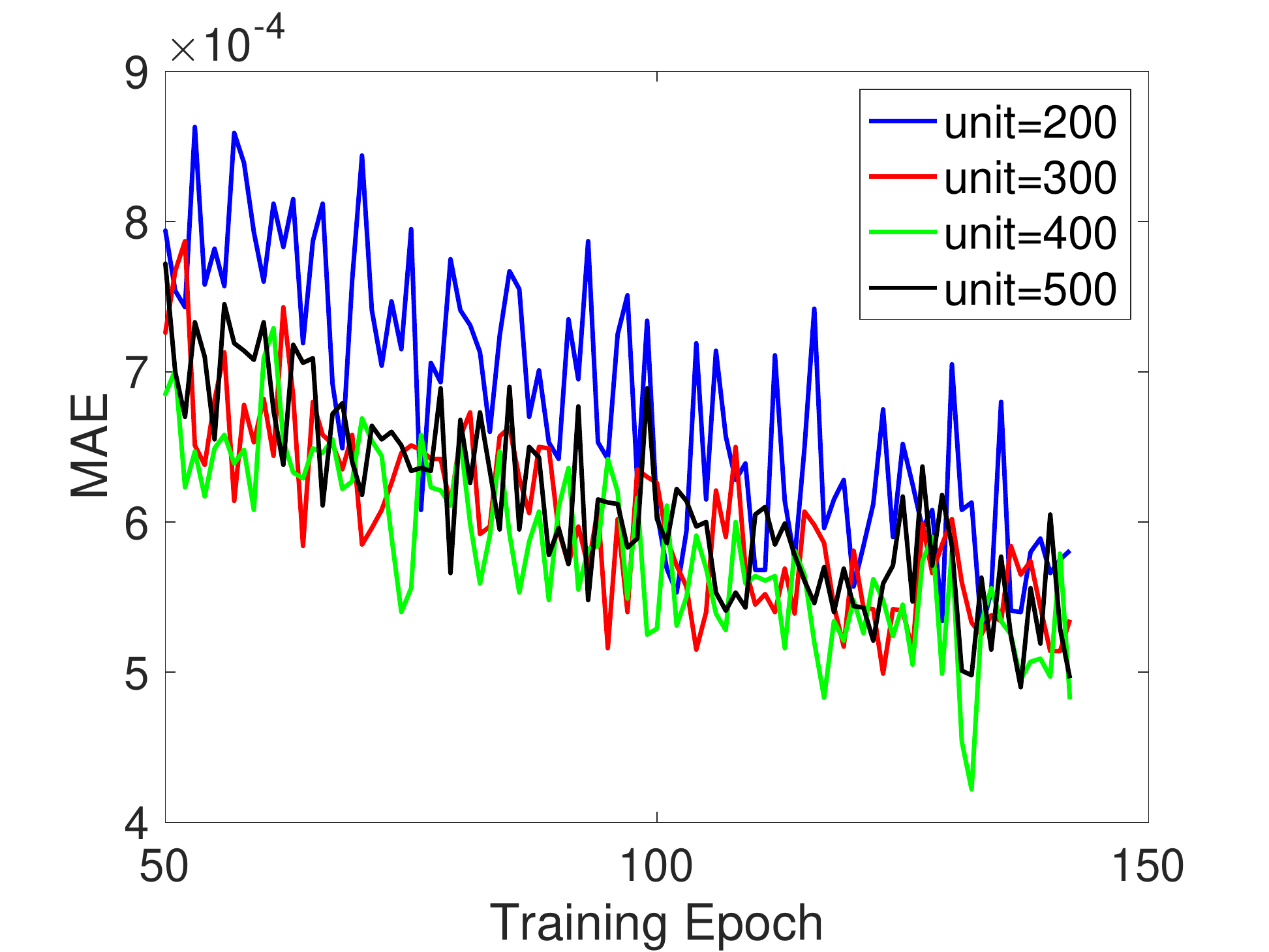}\label{fig-prob1d}} 
    \subfigure[]{\includegraphics[width=0.3\textwidth]{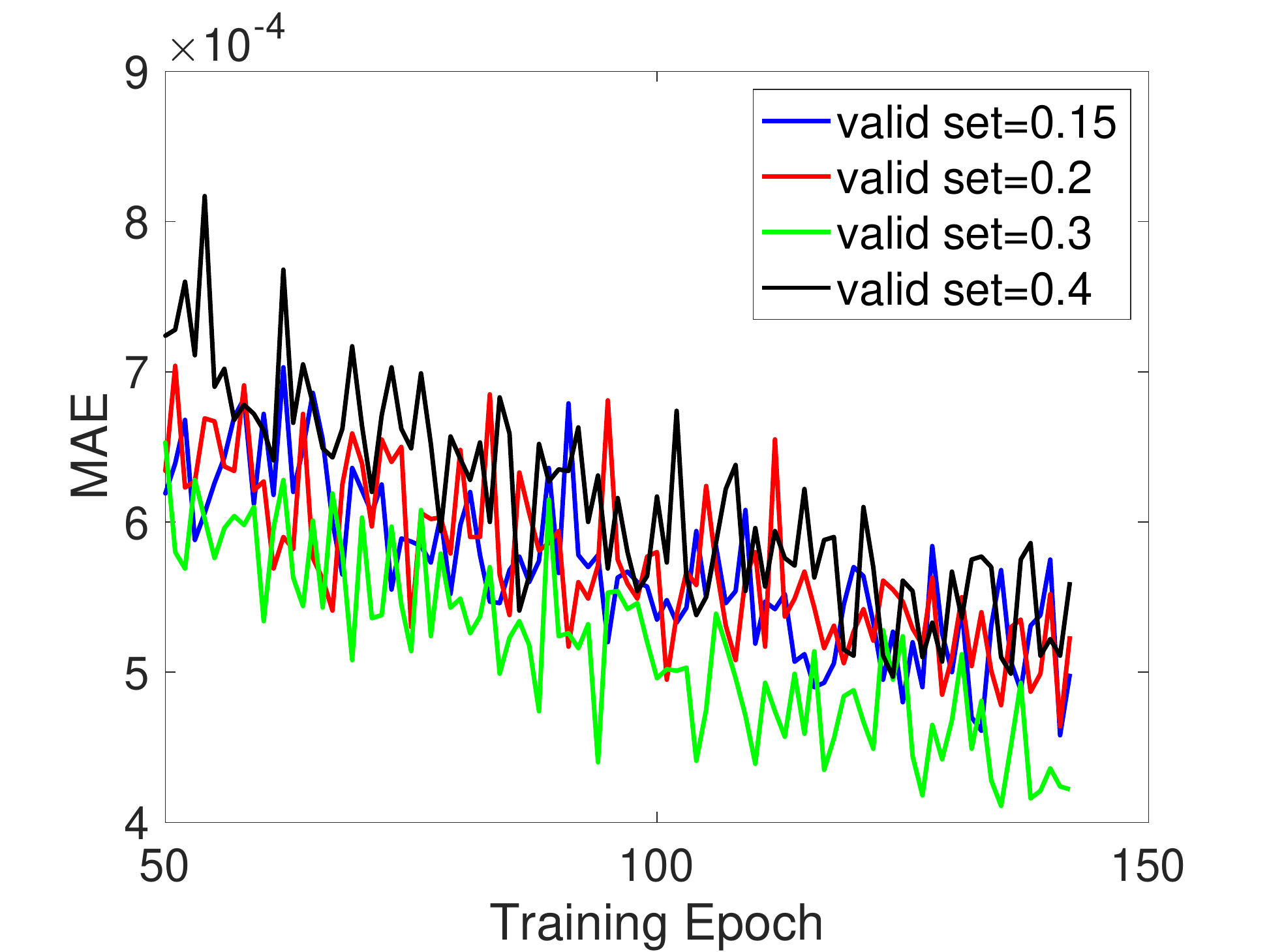}\label{fig-prob1e}} 
    \caption{Results of hyperparameter tuning in Part 2 (Section \ref{sec : problem2}): (a) LeakyReLU Slope, (b) Batch Size, 
    (c) Number of Epoch, and (d) Number of Unit, and (e) Validation Set Size}
    \label{Fig: HyperParameterTuning_Problem1}
\end{figure}

Loading path, history of the loading path, and material properties are considered as input, 
while the Cauchy stress tensors are considered as an output for supervised learning. 
The number of internal units of the LSTM, 
the slope of leaky Relu activation function on negative area, 
the batch size number, the number of epochs, 
and the amount of validation sets are considered as the variables associated with the ADAM optimization approach. 
Figure \ref{Fig: HyperParameterTuning_Problem1} demonstrates the hyperparameter tuning 
by measuring mean absolute errors by minimizing the absolute distance 
between the prediction of Deep Neural Networks and J2 plasticity responses.

\begin{figure}[!htb]
	\centering
	\subfigure[]{\includegraphics[trim=2cm 13cm 6cm 6cm, clip=true, width=0.45\textwidth]{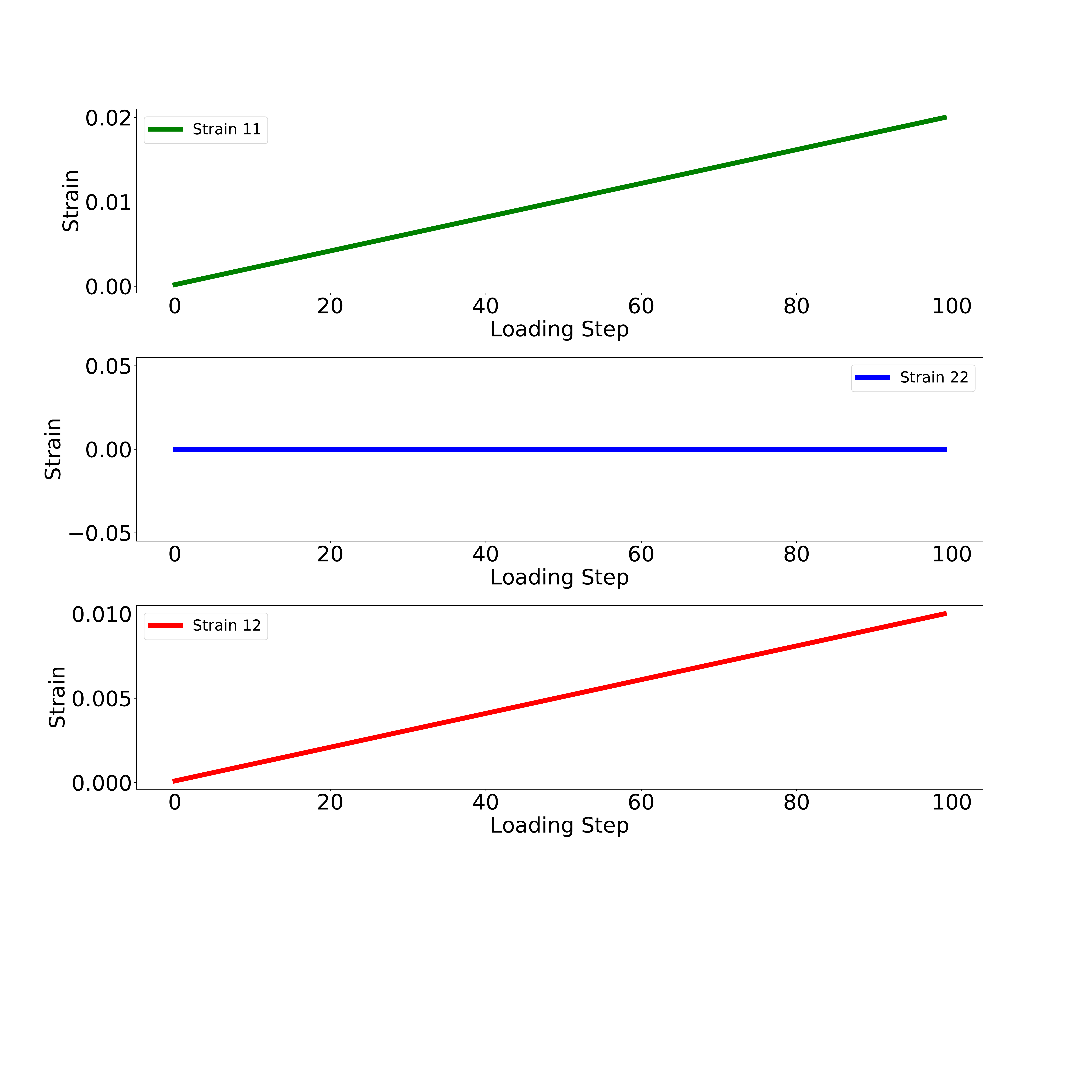}\label{fig-LSTMnoHistoryStrain}} 
	\subfigure[]{\includegraphics[trim=2cm 13cm 6cm 6cm, clip=true, width=0.45\textwidth]{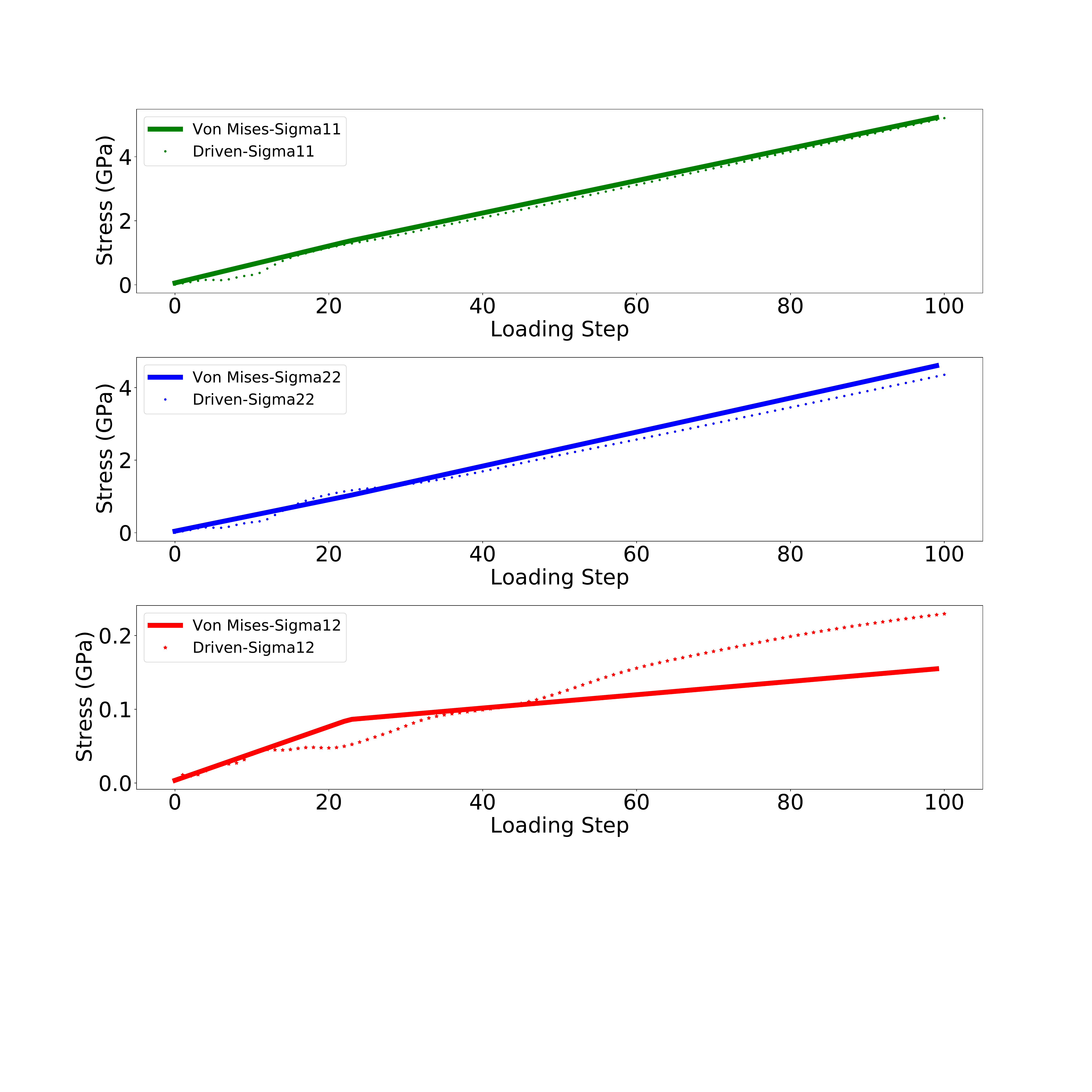}\label{fig-LSTMnoHistoryStress}}
	\caption{Results of the Long-Short Term Memory Unit  (LSTM) without considering the averaged history of strain: 
		(a) Loading path (green: $\epsilon_{xx}$, blue: $\epsilon_{yy}$ and red: $\epsilon_{xy}$), 
		(b) comparison of LSTM prediction and J2 plasticity model (green: $\sigma_{xx}$, blue: $\sigma_{yy}$ and red: $\sigma_{xy}$)} 
	\label{Fig: LSTMnihistory_problem1}
\end{figure}

The basic long-short term memory unit (LSTM) is known not to conserve the dissipation of energy \cite{hoedt2021mc}.
Therefore, this study considers the past averaged strain as an input variable to resolve this issue,
which enhances the inductive bias information over sequence for the LSTM. 
To demonstrate the deficiency of mass conservation of the basic LSTM approach,
a three-stack layered LSTM is adopted for training without feeding the past averaged history. 
The results are shown in Figure \ref{Fig: LSTMnihistory_problem1},
which indicates errors in replicating monotonic loading responses, 
in particular, the off-diagonal component of the stress tensor. 

In this problem, two sets of loading conditions are selected: (1) monotonic, and (2) random loading-unloading. 
To test the model, new loading (Section \ref{sec:DOE})  
and material properties (Table \ref{tab: heterogenity_problem1}) are considered. 
Tables \ref{tab: MaterialPropertiesMonotonicTest_problem1} and \ref{tab: MaterialPropertiesRandomTest_problem1} 
present the three  different material properties under the monotonic and random loading conditions, respectively. 

\begin{table}[!htb]
\caption{Material properties for monotonic loading test (Part 2 - Section \ref{sec : problem2})}
\label{tab: MaterialPropertiesMonotonicTest_problem1}
 \begin{center}
 \begin{tabular}{cccc}
   \hline
   Material properties (GPa) & Test 1  & Test 2 & Test 3\\
   \hline
  $\lambda$ & 373.864 & 296.172 & 295.475\\
  $\mu$ & 34.087 & 33.3992 & 28.1988\\
  $H$ & 37.744 & 31.9616 & 39.22 \\
  $\kappa$ & 0.385419 &  0.252254 & 0.33431\\
  \hline
\end{tabular}
\end{center}
\end{table} 

\begin{table}[h!]
	\caption{Material properties for random loading test (Part 2 - Section \ref{sec : problem2})}
	\label{tab: MaterialPropertiesRandomTest_problem1}
	\begin{center}
		\begin{tabular}{cccc}
			\hline
			Material properties (GPa) & Test 1  & Test 2 & Test 3\\
			\hline
			$\lambda$ & 316.167& 259.166 & 253.007\\
			$\mu$ &  33.1004 & 21.0334 & 21.36\\
			$H$ & 21.4008 & 25.9741 & 30.0251 \\
			$\kappa$ & 0.293411 & 0.236702 & 0.390431\\
			\hline
		\end{tabular}
	\end{center}
\end{table}  

\begin{figure}[h!]
	\centering
	\subfigure[]{\includegraphics[trim=2cm 13cm 6cm 6cm, clip=true,width=0.45\textwidth]{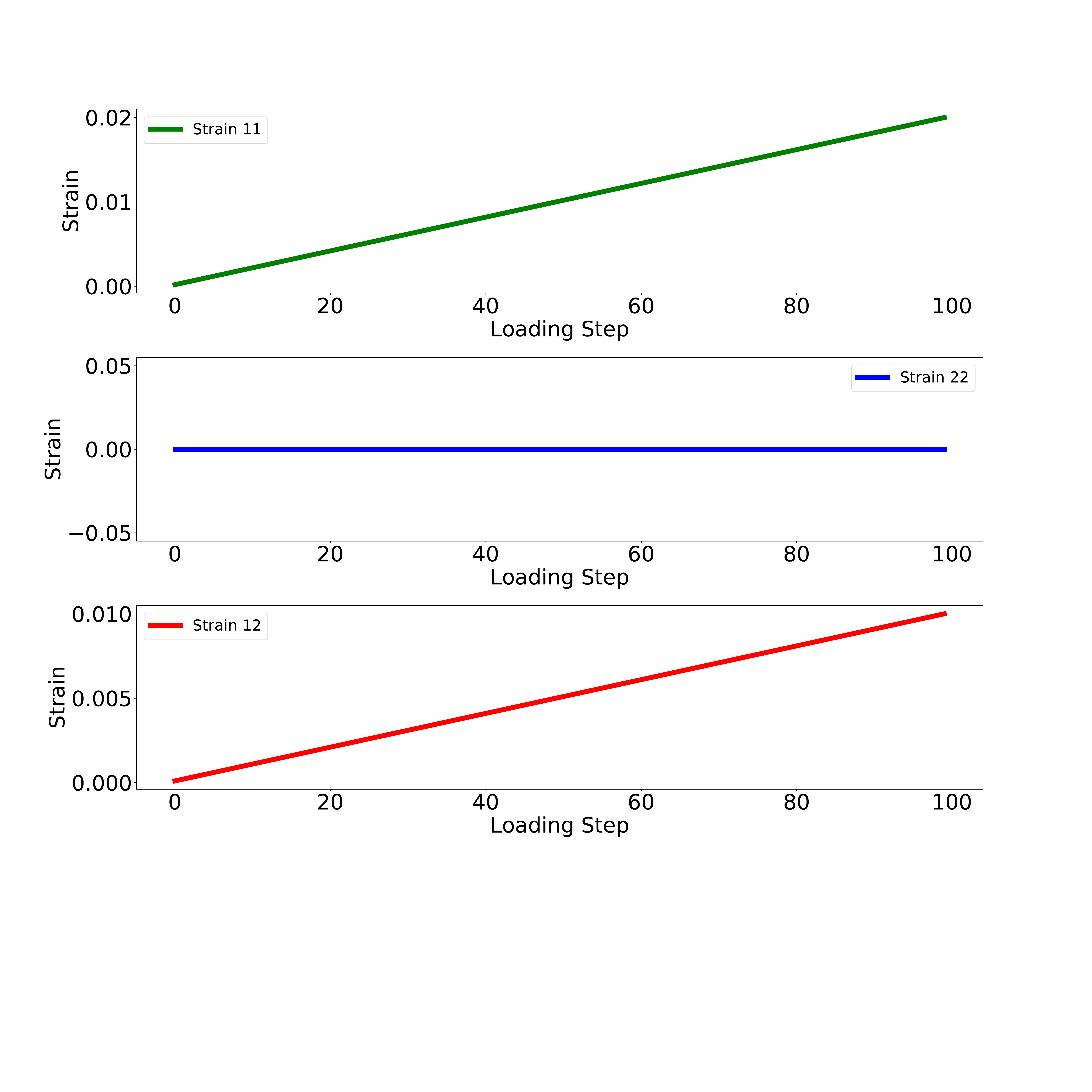}\label{fig-monoprob1a}} 
	\subfigure[]{\includegraphics[trim=2cm 13cm 6cm 6cm, clip=true,width=0.45\textwidth]{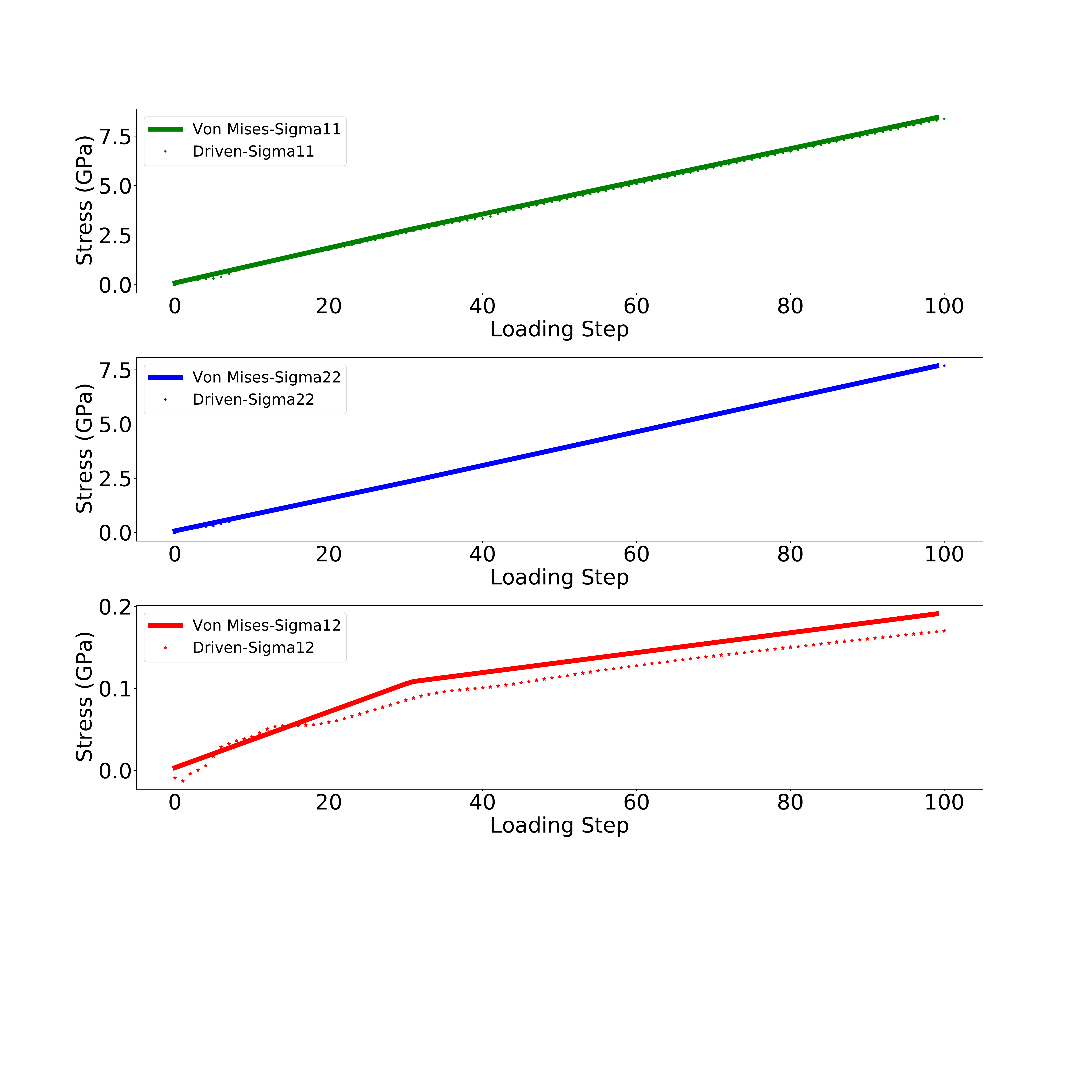}\label{fig-monoprob1b}} 
	\subfigure[]{\includegraphics[trim=2cm 13cm 6cm 6cm, clip=true,width=0.45\textwidth]{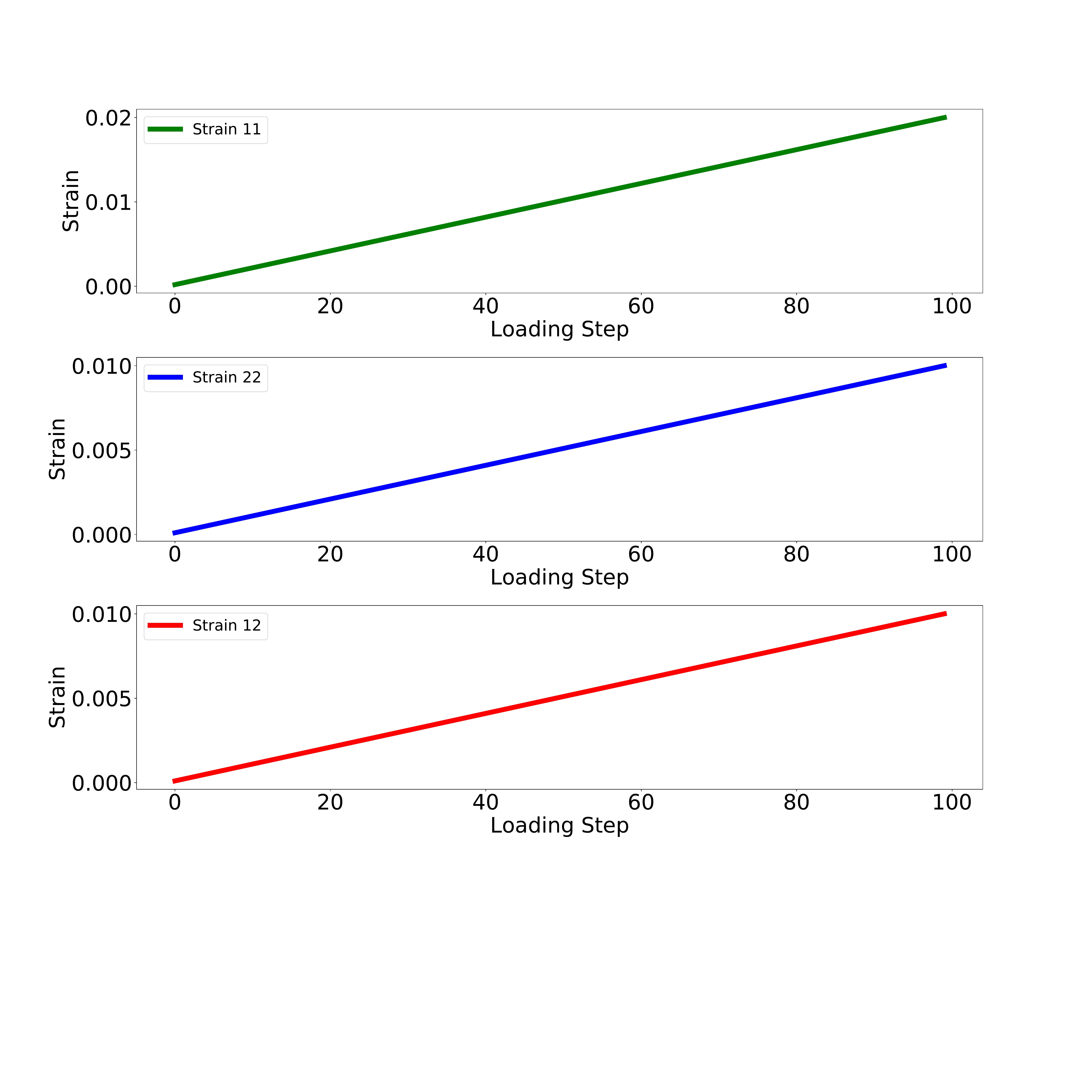}\label{fig-monoprob1e}}     
	\subfigure[]{\includegraphics[trim=2cm 13cm 6cm 6cm, clip=true,width=0.45\textwidth]{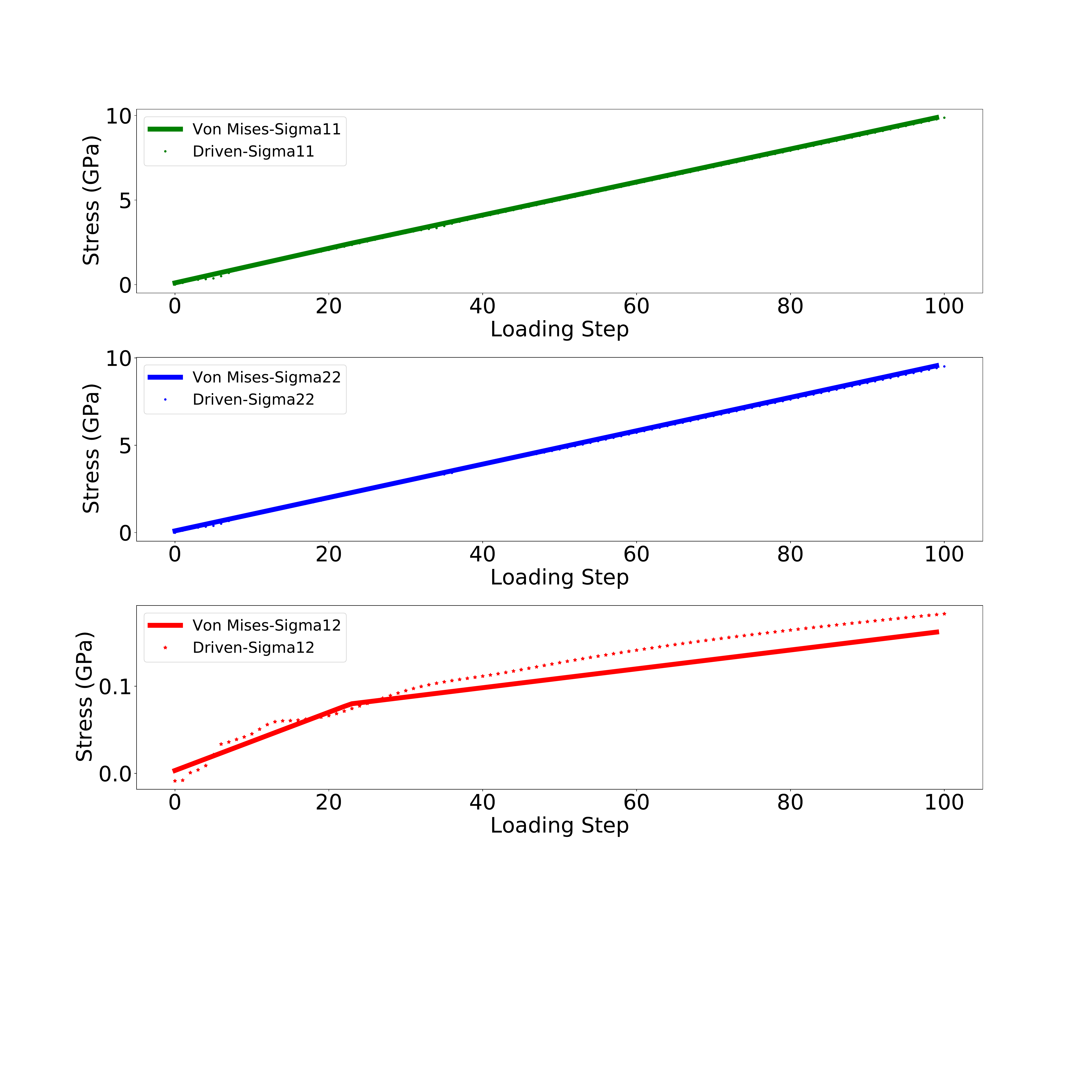}\label{fig-monoprob1a}} 
	\subfigure[]{\includegraphics[trim=2cm 13cm 6cm 6cm, clip=true,width=0.45\textwidth]{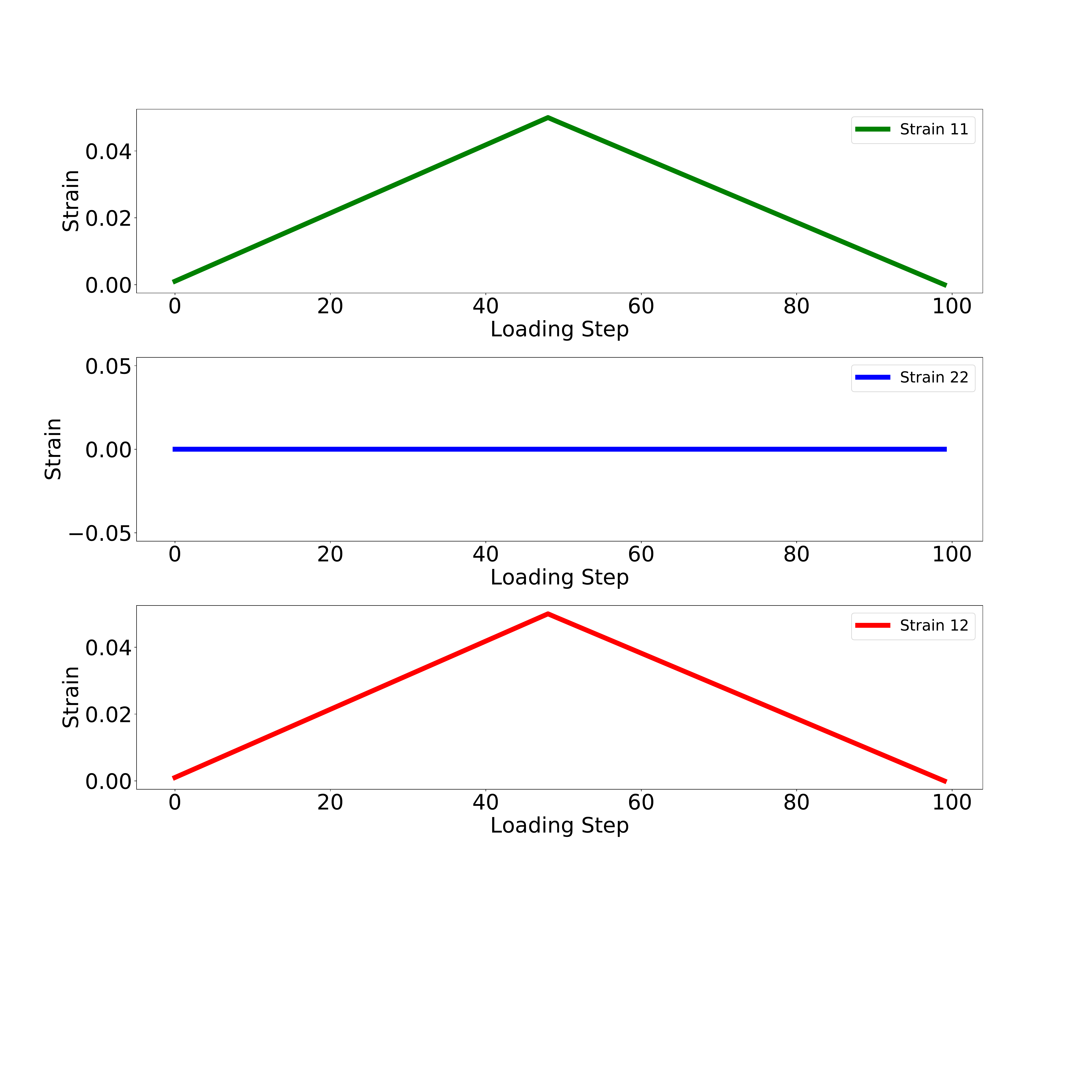}\label{fig-monoprob1b}} 
	\subfigure[]{\includegraphics[trim=2cm 13cm 6cm 6cm, clip=true,width=0.45\textwidth]{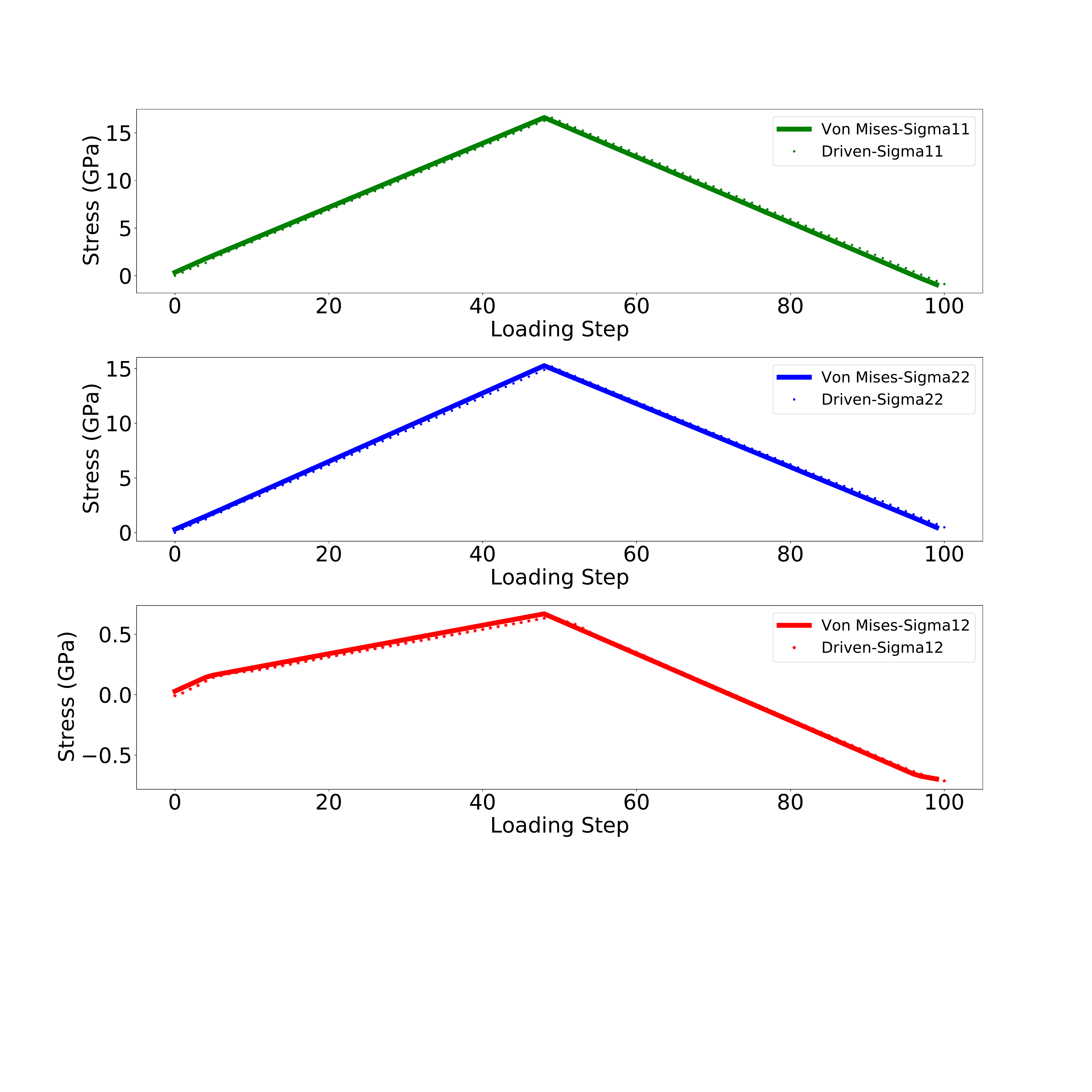}\label{fig-monoprob1c}} 
	\caption{Results of capturing the path-dependent responses associated with J2 plasticity considering material heterogeneity: Data-driven model vs.
		constitutive law under the monotonic loading (Part 2 - Section \ref{sec : problem2})}
	\label{Fig: MonotonicTest_Problem1}
\end{figure}

\begin{figure}[h!]
	\centering
	\subfigure[]{\includegraphics[trim=2cm 13cm 6cm 6cm, clip=true,width=0.45\textwidth]{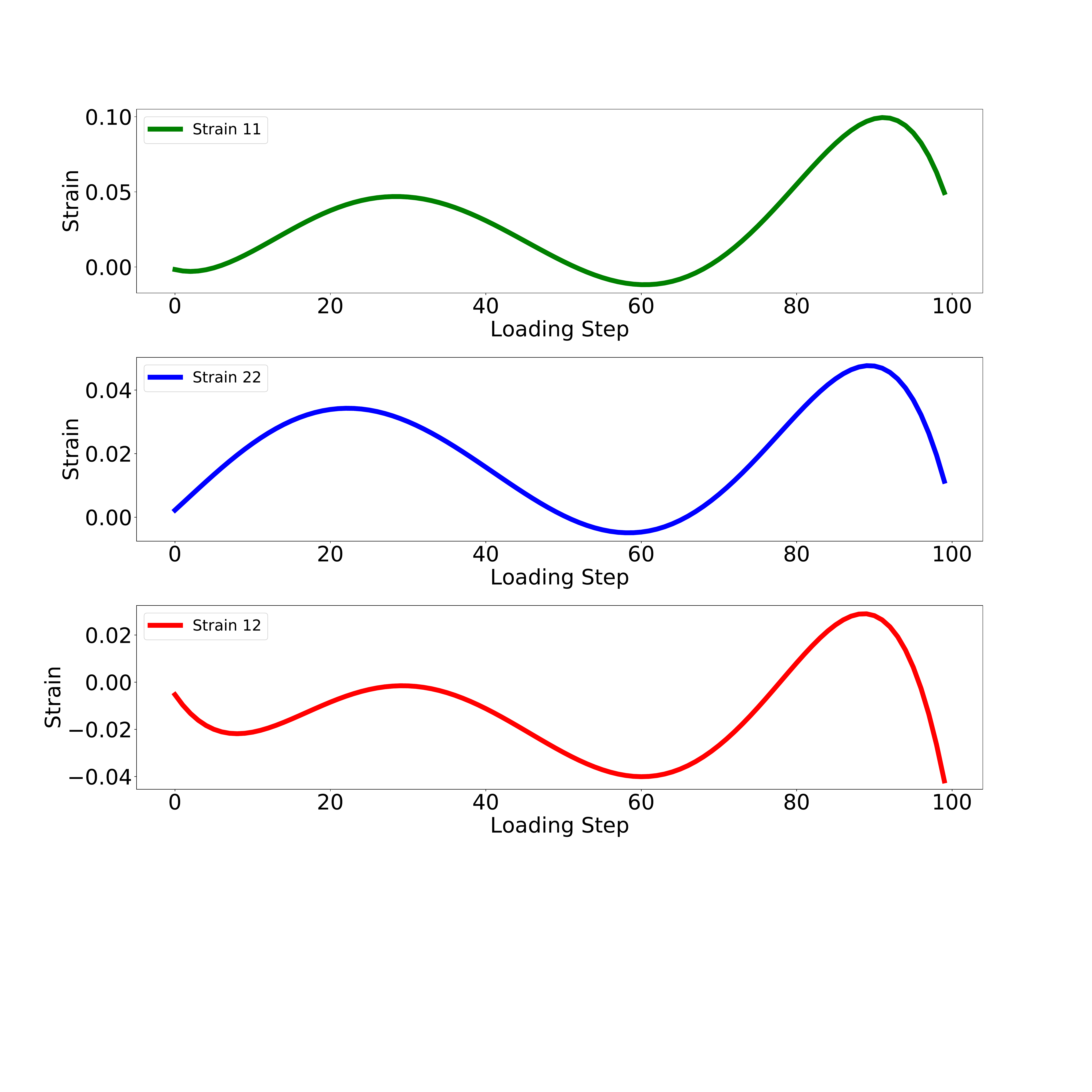}\label{fig-randprob1d}} 
	\subfigure[]{\includegraphics[trim=2cm 13cm 6cm 6cm, clip=true,width=0.45\textwidth]{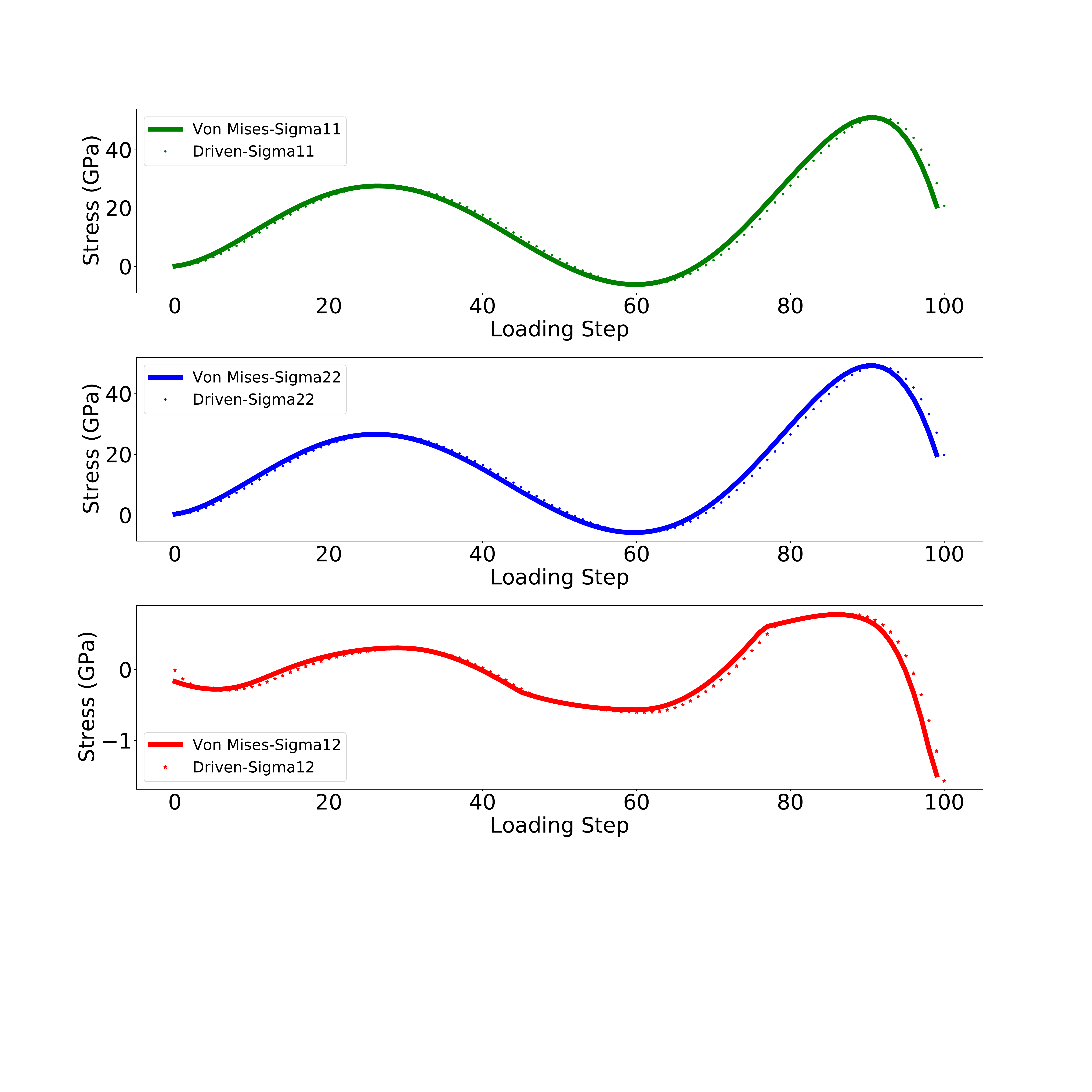}\label{fig-randprob1e}}     
	\subfigure[]{\includegraphics[trim=2cm 13cm 6cm 6cm, clip=true,width=0.45\textwidth]{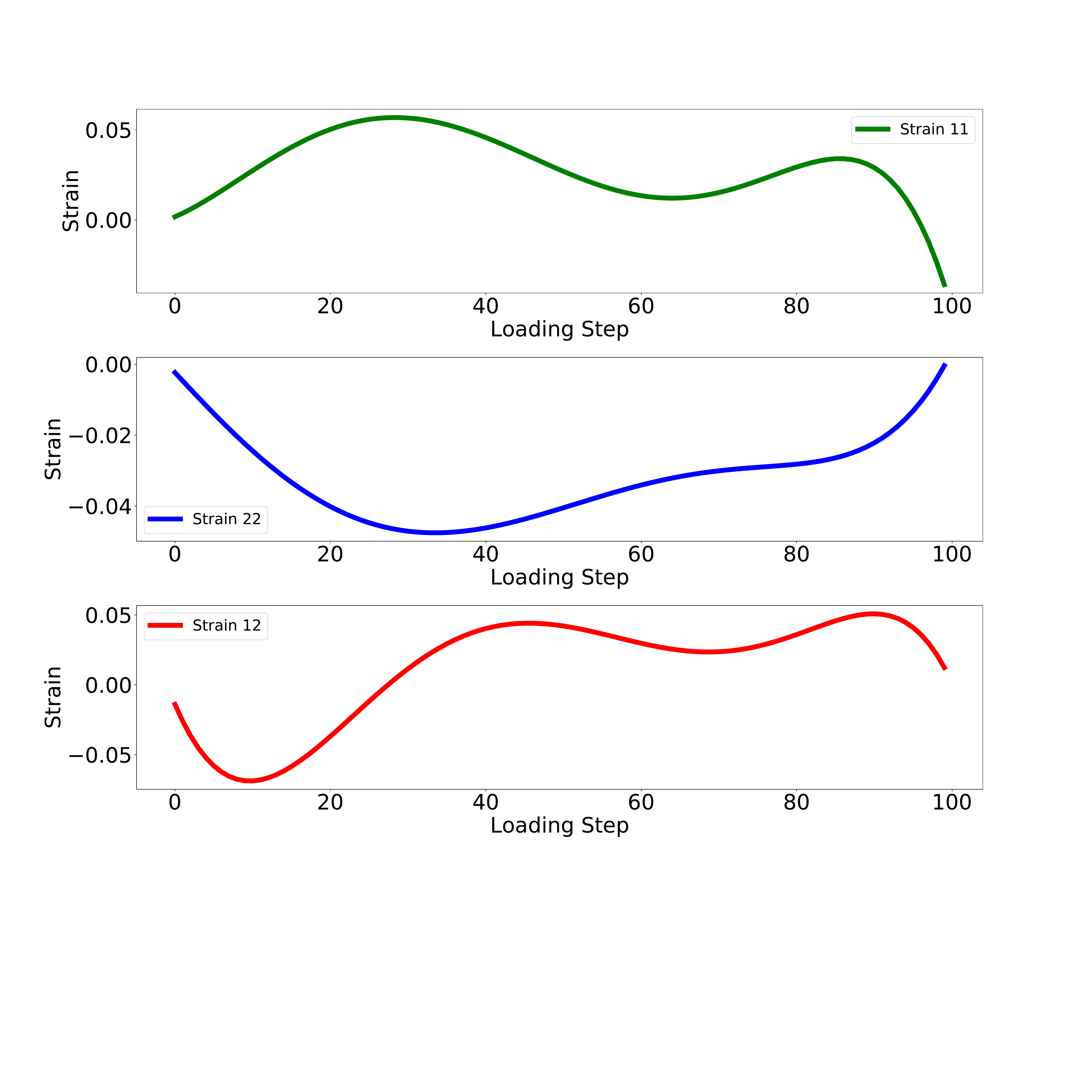}\label{fig-randprob1a}} 
	\subfigure[]{\includegraphics[trim=2cm 13cm 6cm 6cm, clip=true,width=0.45\textwidth]{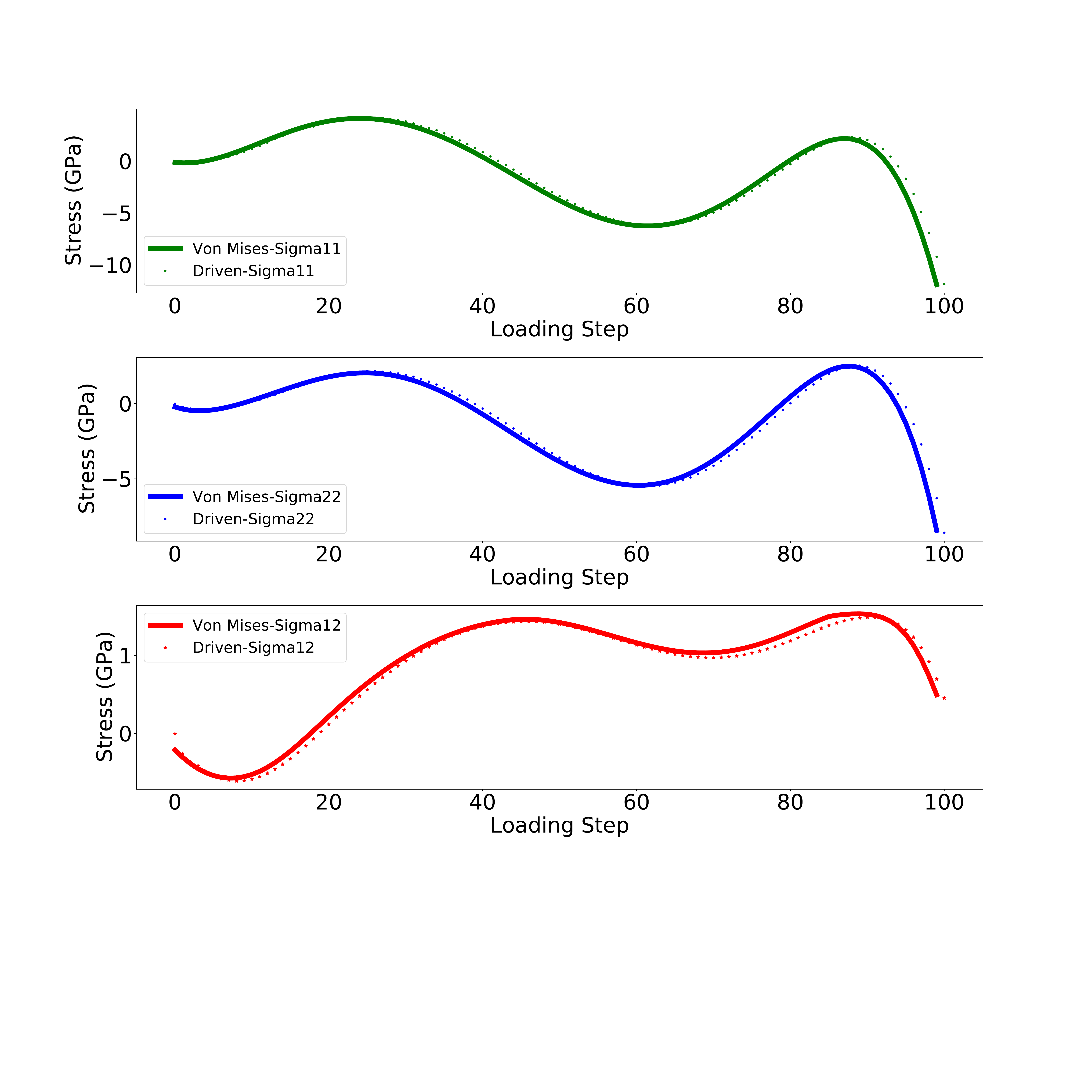}\label{fig-randprob1b}} 
	\subfigure[]{\includegraphics[trim=2cm 13cm 6cm 6cm, clip=true,width=0.45\textwidth]{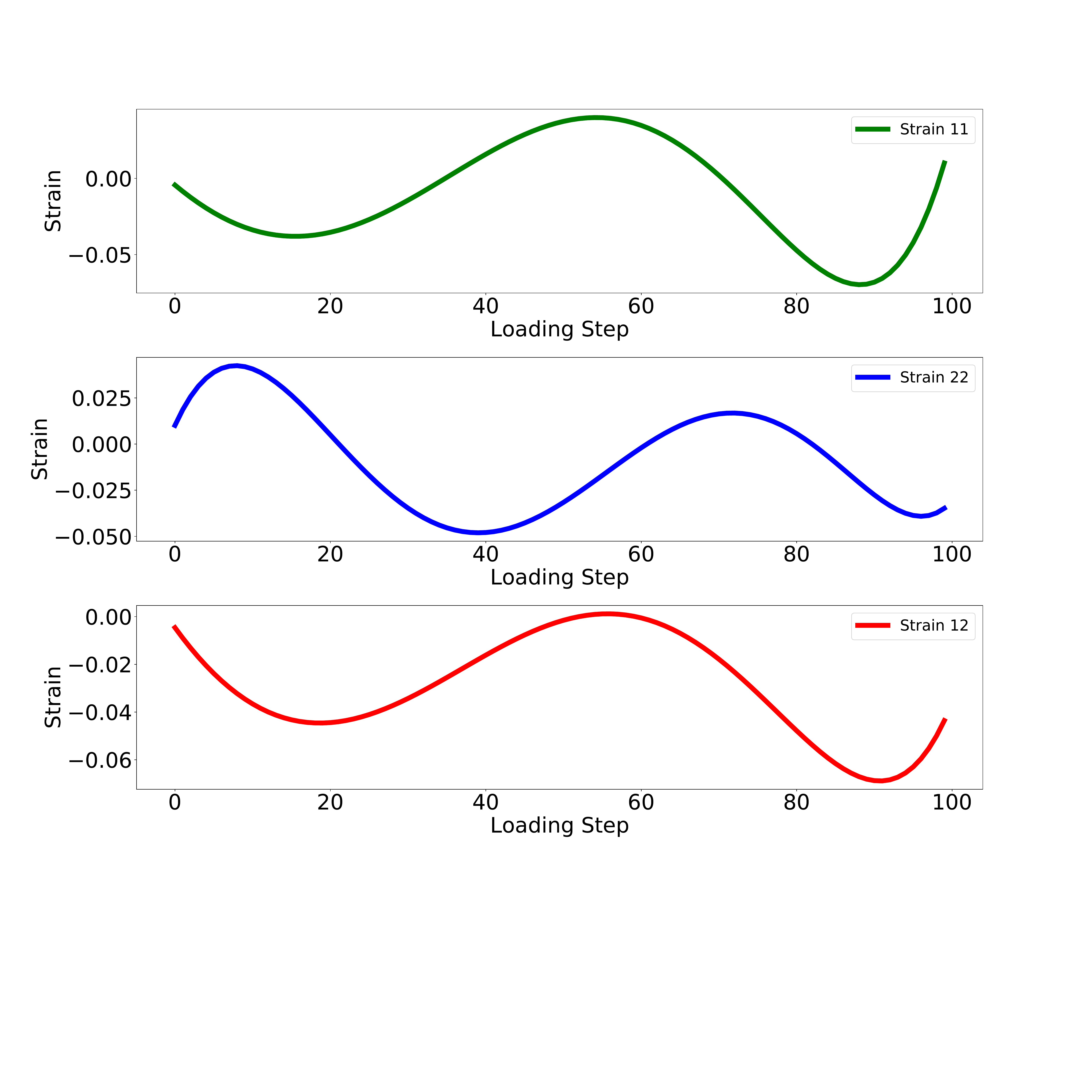}\label{fig-randprob1c}} 
	\subfigure[]{\includegraphics[trim=2cm 13cm 6cm 6cm, clip=true,width=0.45\textwidth]{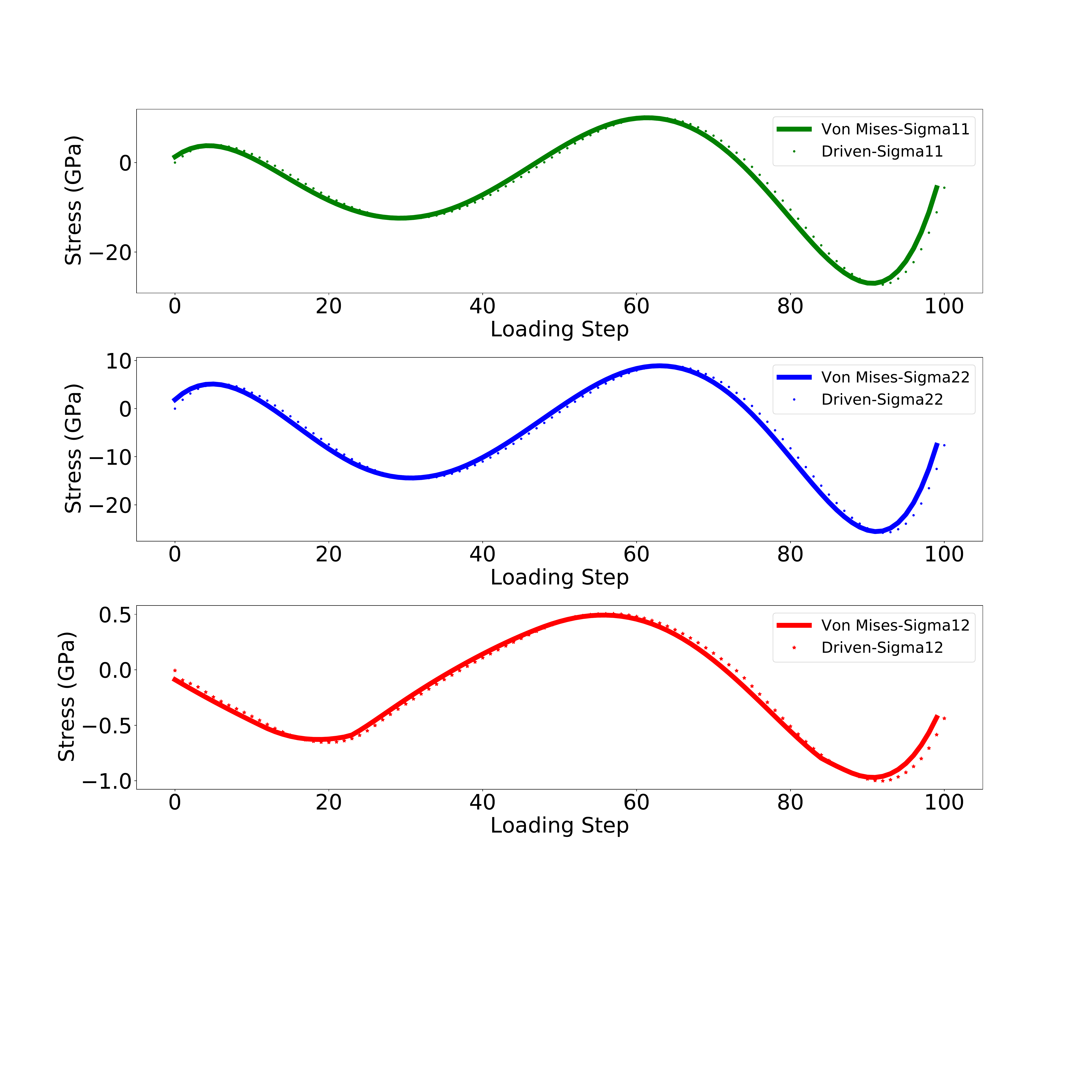}\label{fig-randprob1d}} 
	
	\caption{Results of capturing the path-dependent responses associated with J2 plasticity considering material heterogeneity: Data-driven model vs.
		constitutive law under the random loading (Part 2 - Section \ref{sec : problem2})}
	\label{Fig: RandomTest_Problem1}
\end{figure}

Figure \ref{Fig: MonotonicTest_Problem1} (b,d,f), 
and Figure \ref{Fig: RandomTest_Problem1} (b,d,f) demonstrate the results of comparison 
between the response of deep recurrent neural network and hardening 
J2 plasticity model. 

\begin{figure}[h!]
    \centering
    \subfigure[]{\includegraphics[width=0.3\textwidth]{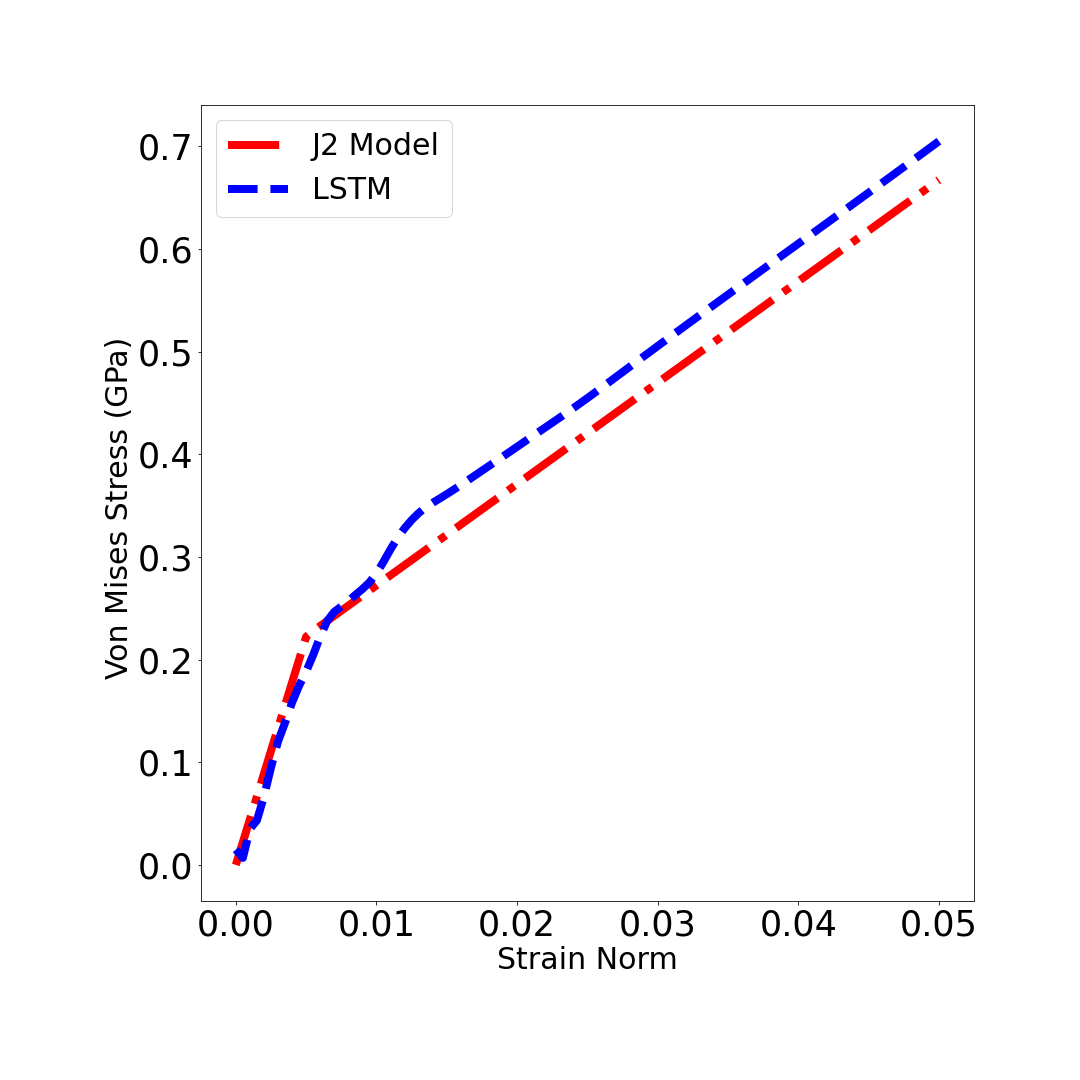}\label{fig-vonmises1_problem1}} 
    \subfigure[]{\includegraphics[width=0.3\textwidth]{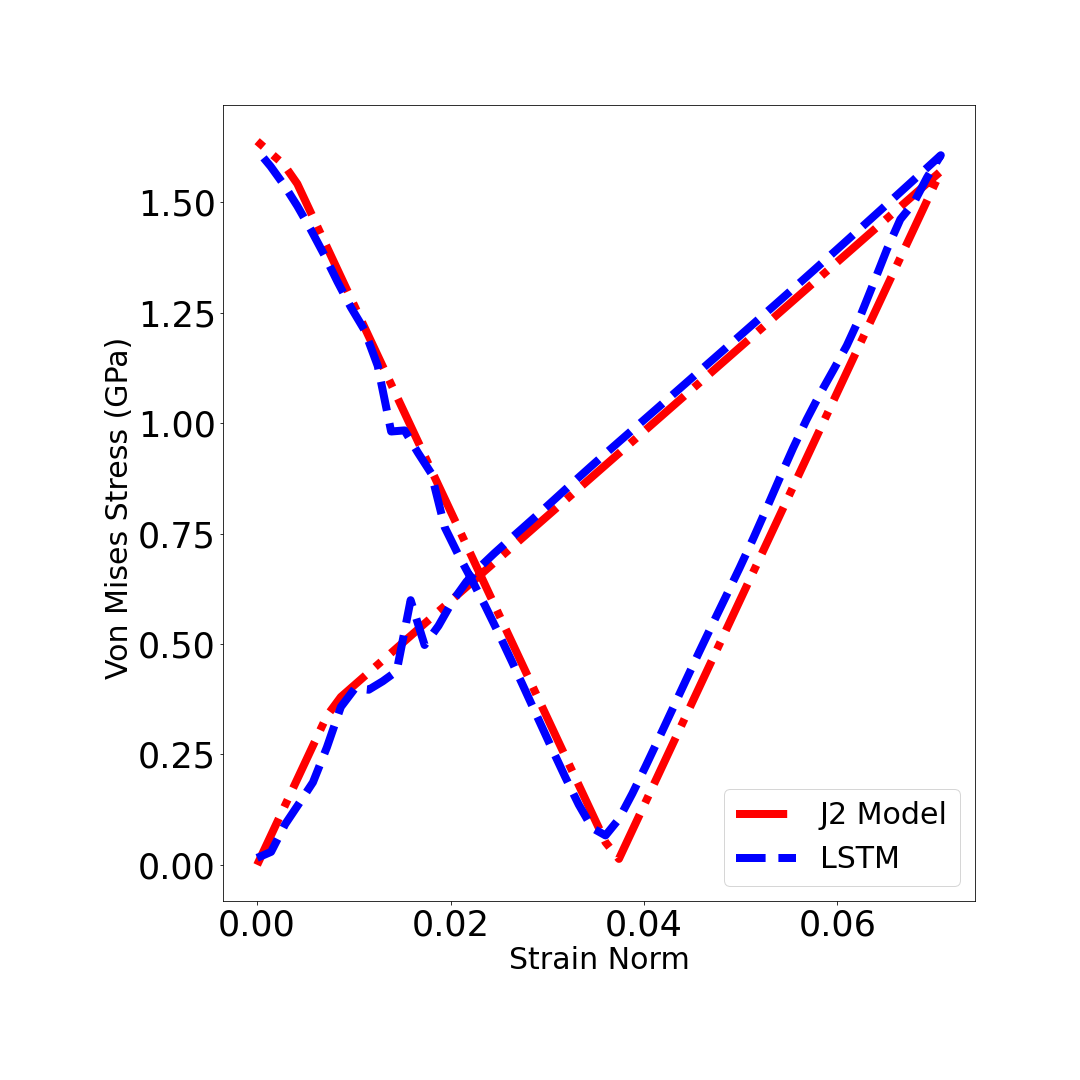}\label{fig-vonmises2_problem1}} 
    \subfigure[]{\includegraphics[width=0.3\textwidth]{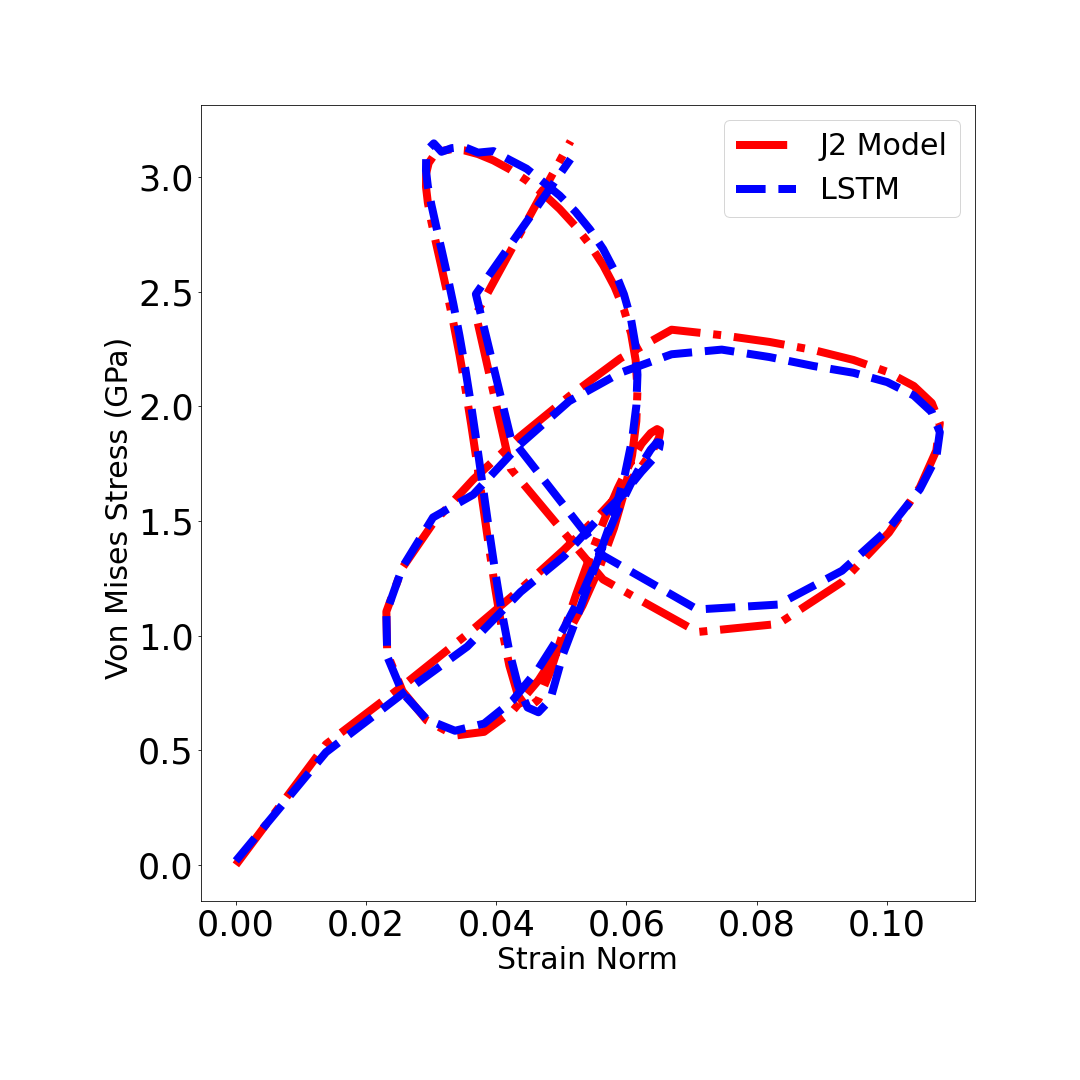}\label{fig-vonmises3_problem1}} 
    \caption{Comparison of overall responses with J2 plasticity and the LSTM based model considering material heterogeneity: (a) Monotonic Loading (Material properties in Table \ref{tab: MaterialPropertiesMonotonicTest_problem1}), (b) Monotonic Loading-Unloading (Material properties in Table \ref{tab: MaterialPropertiesMonotonicTest_problem1}), (c) Random Loading-Unloading (Material properties in Table \ref{tab: MaterialPropertiesRandomTest_problem1})}
    \label{Fig: vonmisesstress_problem1}
\end{figure}

Figure \ref{Fig: vonmisesstress_problem1} 
depicts the J2 plasticity stress for above test.  
The loading paths are depicted with 3 components of 
strain ($\epsilon_{xx}$, $\epsilon_{yy}$ and $\epsilon_{xy}$), and the prediction 
is depicted with 3 components of stress ($\sigma_{xx}$, $\sigma_{yy}$ and $\sigma_{xy}$). 
The proposed strategy shows that a single deep neural network architecture can capture heterogeneous path-dependent 
behavior with an error of 2 percent. 

\subsection{Part 3: Learning path-dependent behavior of  anisotropic microstructures}
\label{Sec : Problem3}
\begin{figure}[h!]
	\centering
	\subfigure[]{\includegraphics[width=0.205\textwidth]{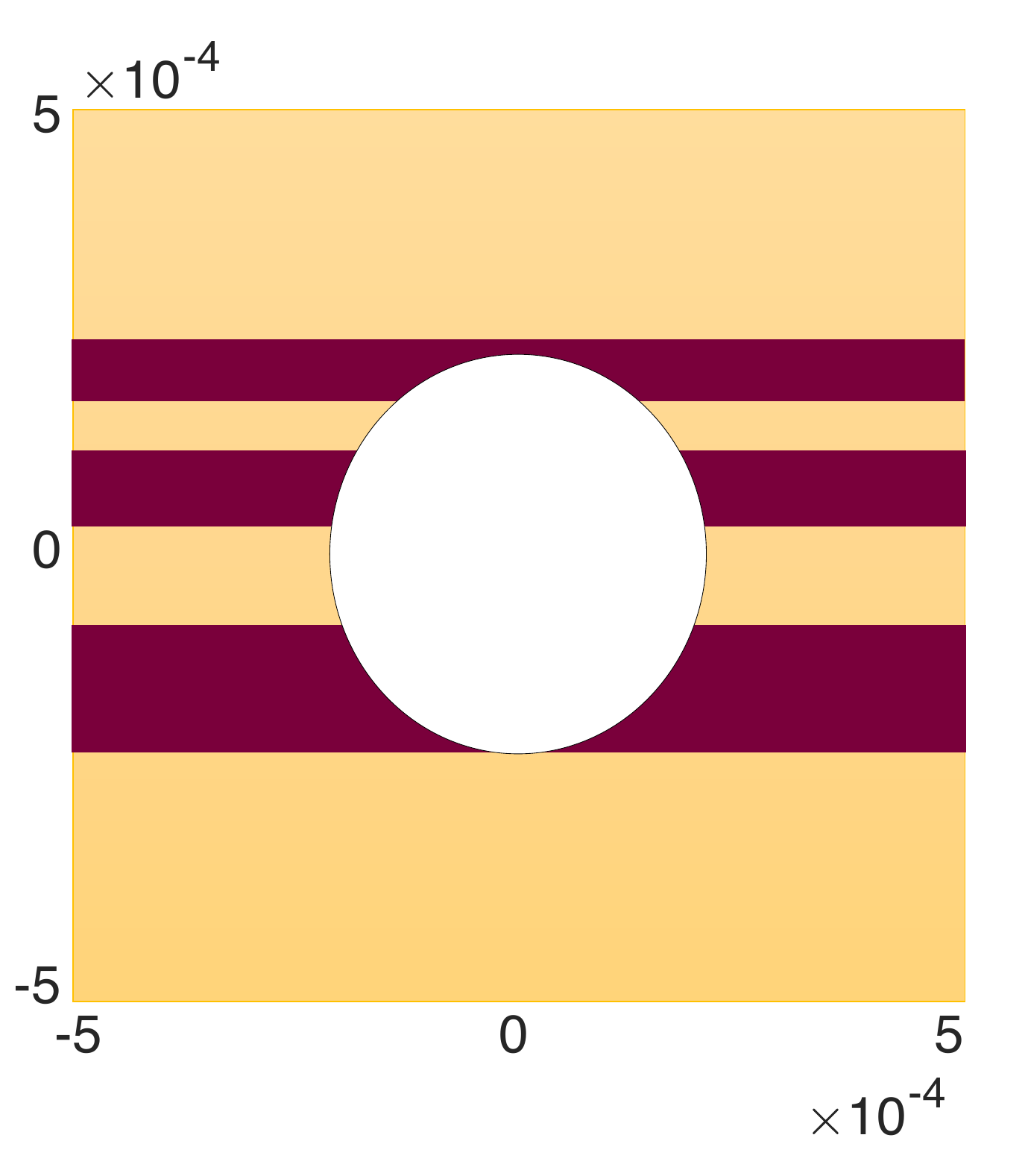}\label{fig-anisomicro1}} 
	\subfigure[]{\includegraphics[width=0.3\textwidth]{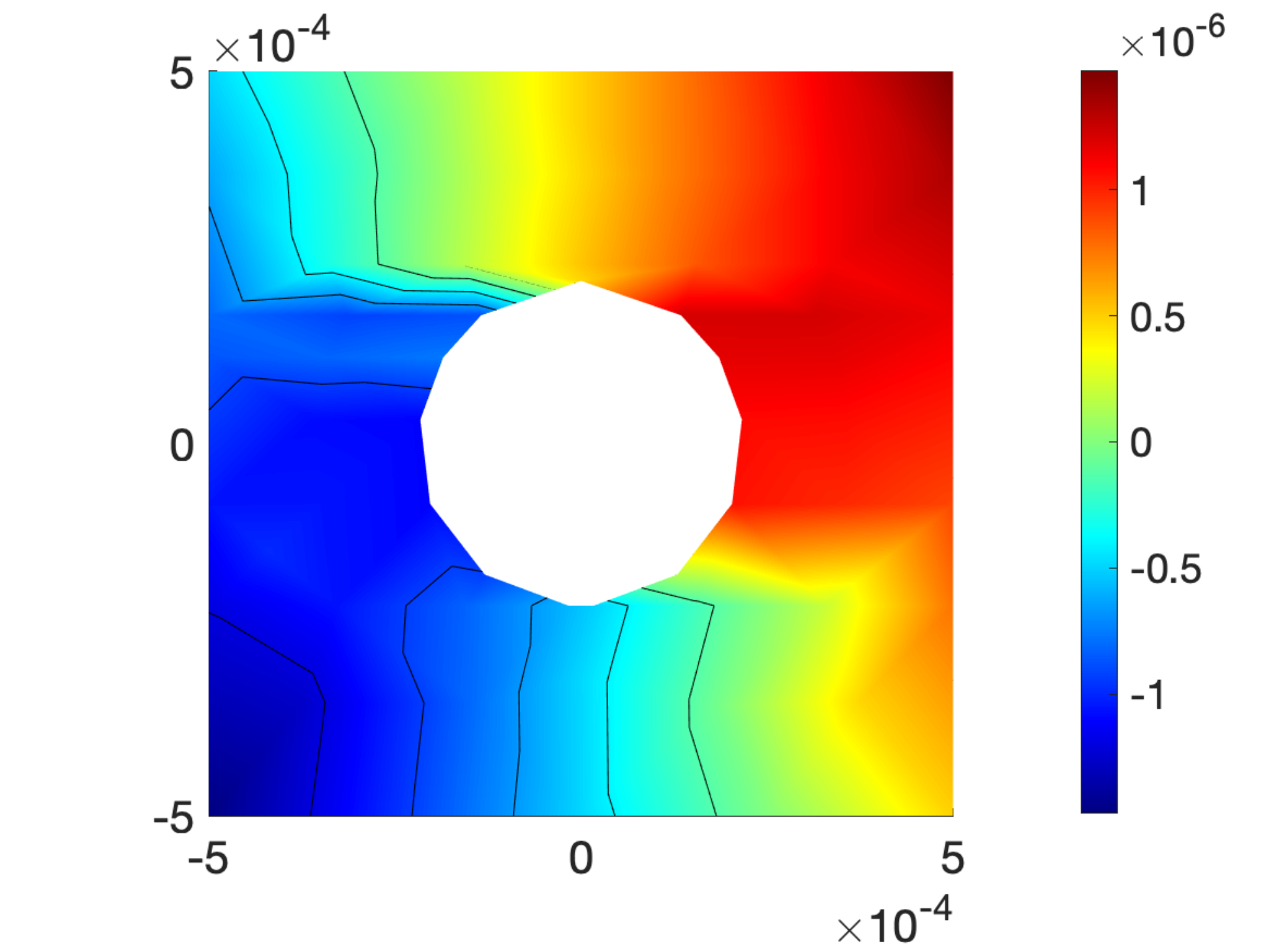}\label{fig-anisomicro2}} 
	\subfigure[]{\includegraphics[width=0.3\textwidth]{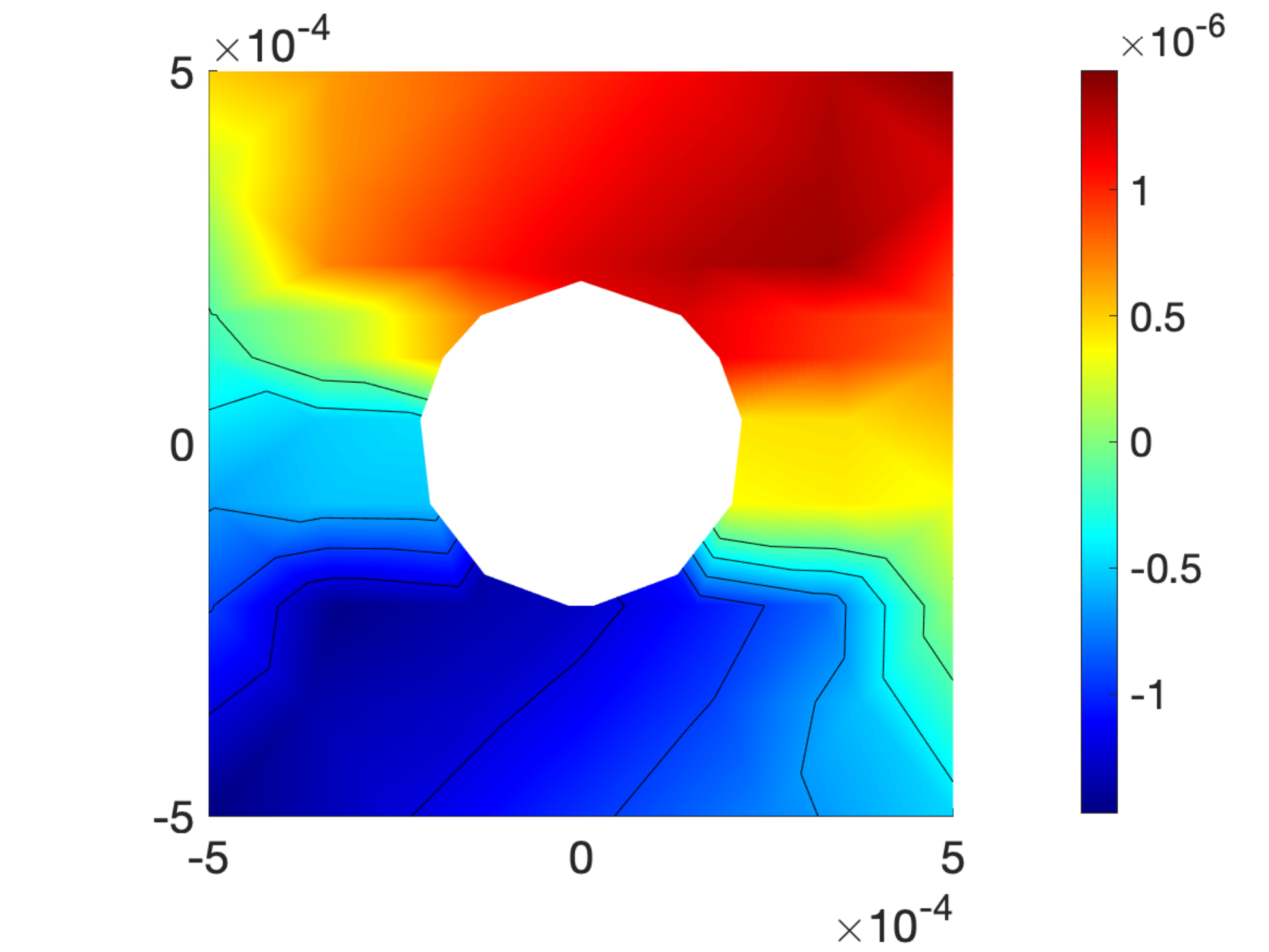}\label{fig-anisomicro3}} 
	\caption{Schematic and simulation results of a sample microstructure under the monotonic Loading, (a) Anisotropic microstructure, (b) X-Displacement, (c) Y-Displacement - the units are in m}1
	\label{Fig: anisosample disp_problem3}
\end{figure}

\begin{figure}[h!]
	\centering
	\includegraphics[width=.4\textwidth]{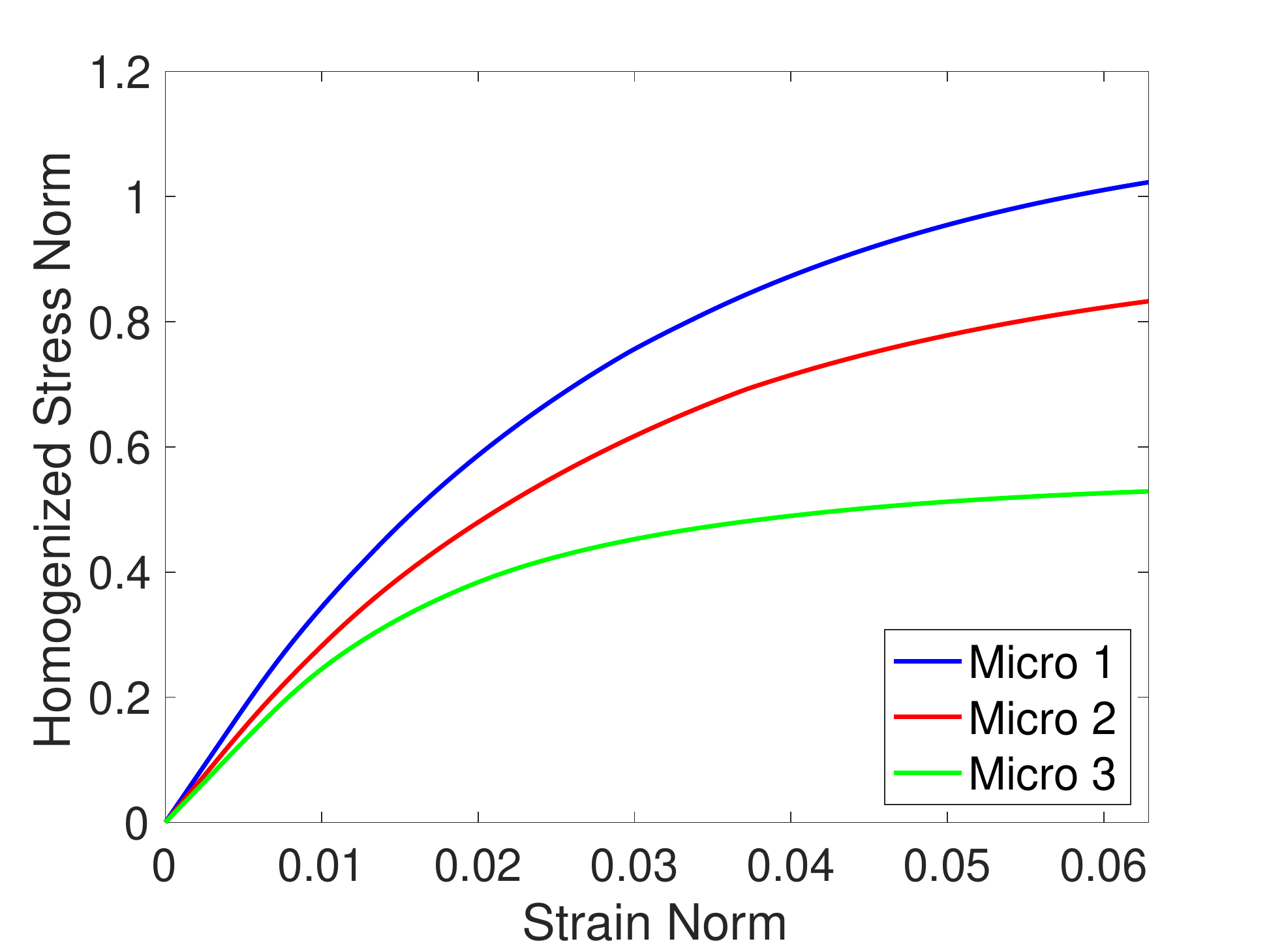}
	\caption{Comparison between the homogenized responses of three anisotropic microstructures in Table \ref{tab: descriptor_problem2}} 
	\label{Fig: Comparisono_anisomicro_problem3}
\end{figure}

Finally, the proposed  single LSTM network is applied for capturing the path-dependent responses
of heterogeneous anisotropic microstructures. 
This problem is designed and investigated as follows:
(1) 8,000 transversely isotropic microstructures are randomly generated (Section \ref{sec:DOE}), 
(2) Three descriptors are defined to identify each microstructure's distinctive features,
that are the relative location, thickness, and volumetric fraction of layers 
(refer to Table \ref{tab: descriptor_aniso_section3}),
(3) Randomly generated loading paths (Section \ref{sec:DOE}) are applied to each microstructure, 
(4) Homogenized responses of 8,000 microstructures under each loading path are collected
using the FE$^2$ framework (Section \ref{sec : FEM2}). 
Figure \ref{Fig: anisosample disp_problem3} presents a sample simulation to show the microstructural attribute 
and the displacement distribution under the monotonic loading condition. 
In addition, Figure \ref{Fig: Comparisono_anisomicro_problem3} demonstrates the homogenized behavior of three
anisotropic microstructures under monotonic loading and as it appears the distribution 
of different layers results in distinct behaviors.

After the data collection, input data is prepared by concatenating 
strain paths (3 components of each strain tensor), geometric descriptors, and average strain;
output data is prepared via the stress path (3 components of each stress tensor). 
Similar to Part 2 in the preceding section, 
the single architecture of deep neural networks is considered using 3 stacked layers 
of long-short term memory unit (LSTM), which is described in Table \ref{tab: ArchitectureofNetwork_problem2}.
The model is trained after running a simple hyperparameter tuning on the slope of Leaky 
ReLu activation function, number of epochs, the number of LSTM units, batch size, 
and validation set ratio. 
Adam optimizer via mean absolute error is also used for the training. 

\begin{table}[h!]
	\caption{Architecture of the Deep Long-Short Term Memory (LSTM) Unit for Part 3 (Section \ref{Sec : Problem3})}
	\label{tab: ArchitectureofNetwork_problem2}
	\begin{center}
		\begin{tabular}{ccc}
			\hline
			Layer (Type) & Output Shape  & Activation Function\\
			\hline
			Input & (None, 101, 13) & None  \\
			LSTM & (None,101,300) & tanh \\
			LSTM & (None,101,300) & tanh \\
			LSTM & (None,101,300) & tanh\\
			Time Distributed & (None, 101, 3) & LeakyReLU  \\
			\hline
		\end{tabular}
	\end{center}
\end{table}

Similar to the problem of Part 2 in Section \ref{sec : problem2},
a newly generated random microstructure, not used in the training, is selected
to test the proposed LSTM-based model. 
Table \ref{tab: descriptor_problem2} depicts the descriptors of the newly generated random structure. 

\begin{table}[h!]
	\caption{Descriptors for Anisotropic Microstructure for Part 3 (Section \ref{Sec : Problem3})}
	\label{tab: descriptor_problem2}
	\begin{center}
		\begin{tabular}{cccc}
			\hline
			Descriptor & Micro 1 & Micro 2 & Micro 3  \\
			\hline
			Relative Fraction of Layers & 0.2891  & 0.3576 & 0.4068\\
			Relative Thickness & [0.137, 0.0835, 0.069] & [0.137, 0.084, 0.067] &[0.421, 0.064, 0.065]  \\
			Relative Location & [0.282, 0.532, 0.672] & [0.282, 0.532, 0.672]& [0.0792, 0.631, 0.801]  \\
			\hline
		\end{tabular}
	\end{center}
\end{table}

\begin{figure}[!t]
    \centering
    \subfigure[]{\includegraphics[trim=2cm 13cm 6cm 6cm, clip=true,width=0.45\textwidth]{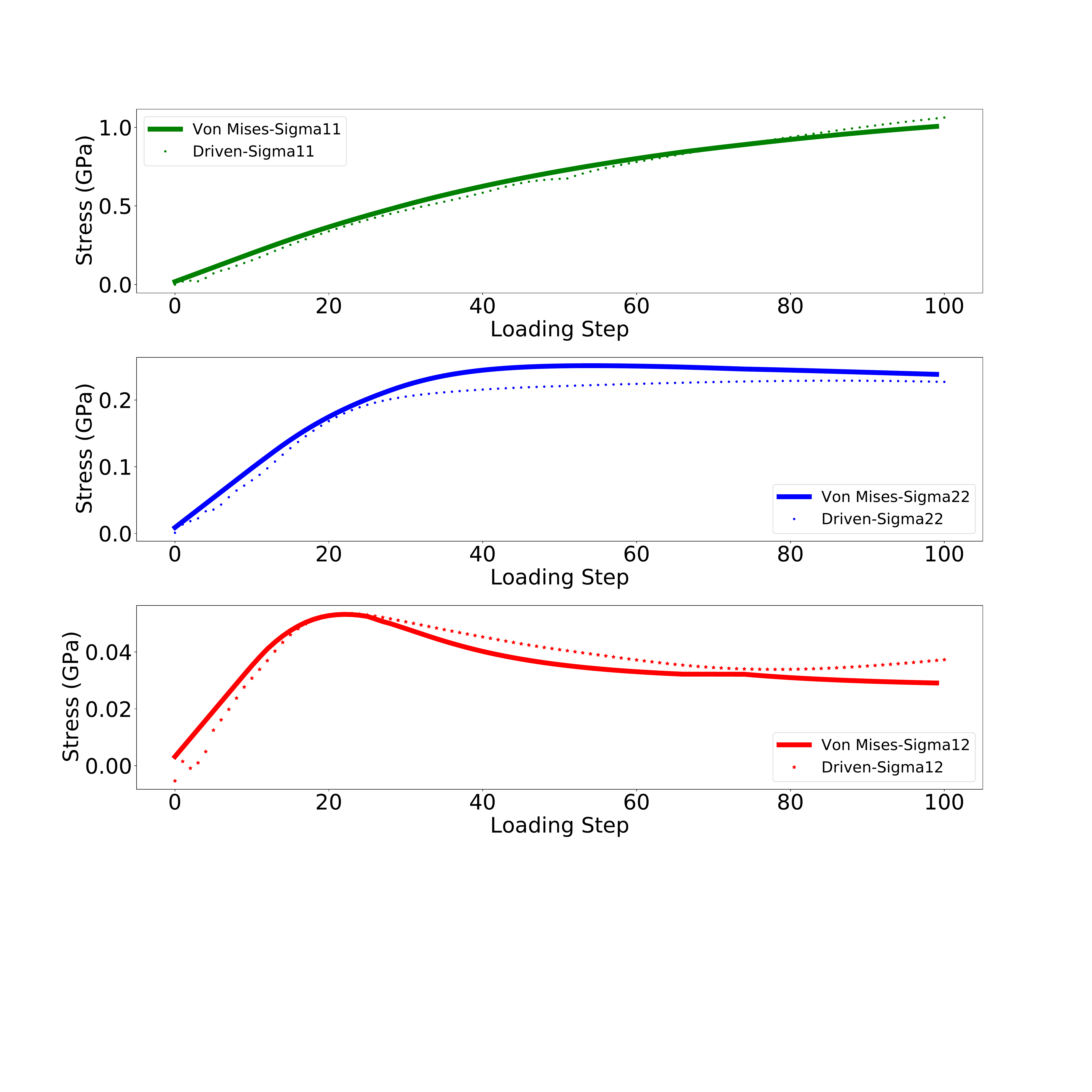}\label{fig-randprob1d}}
    \subfigure[]{\includegraphics[trim=2cm 13cm 6cm 6cm, clip=true,width=0.45\textwidth]{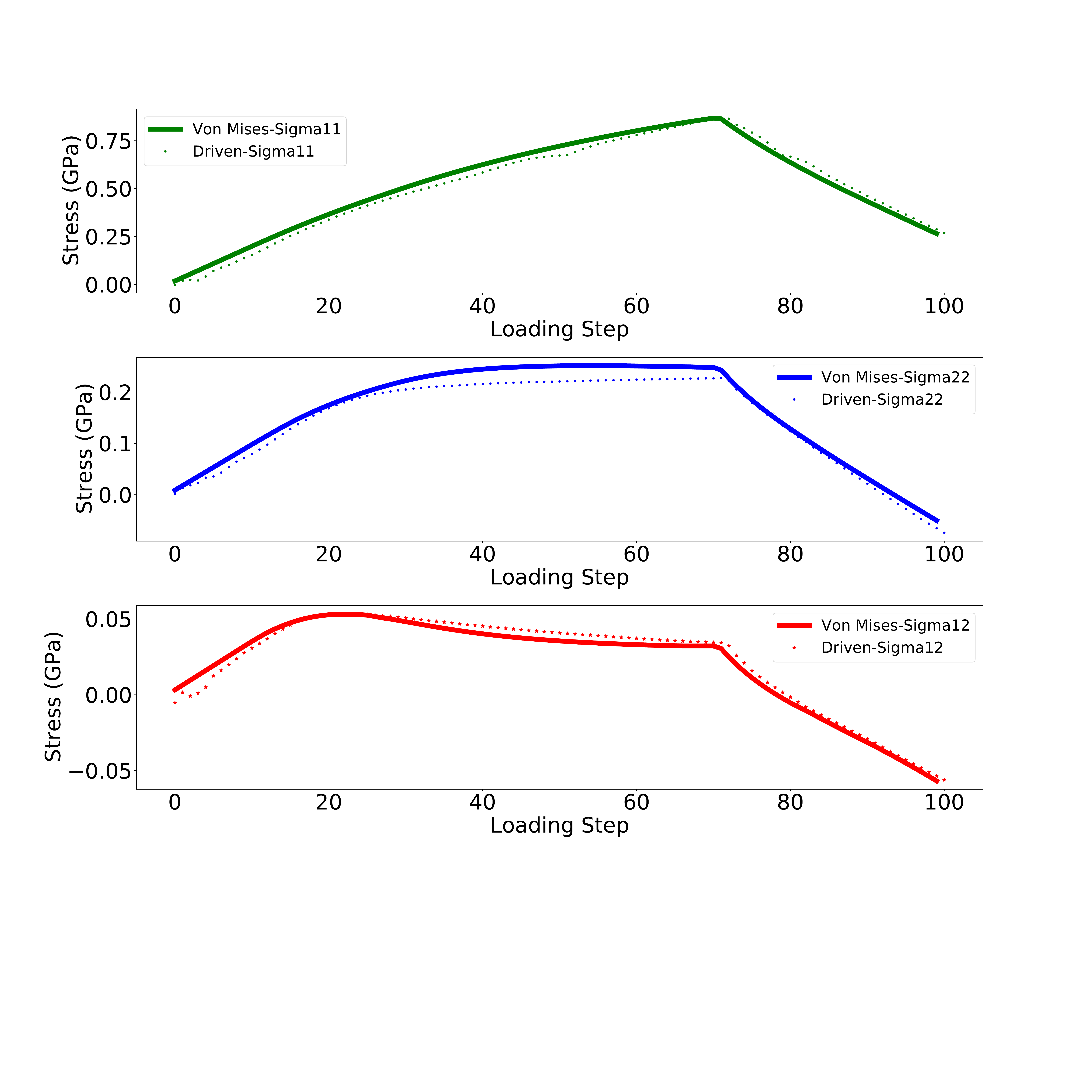}\label{fig-randprob1e}} 
    \subfigure[]{\includegraphics[trim=2cm 13cm 6cm 6cm, clip=true,width=0.45\textwidth]{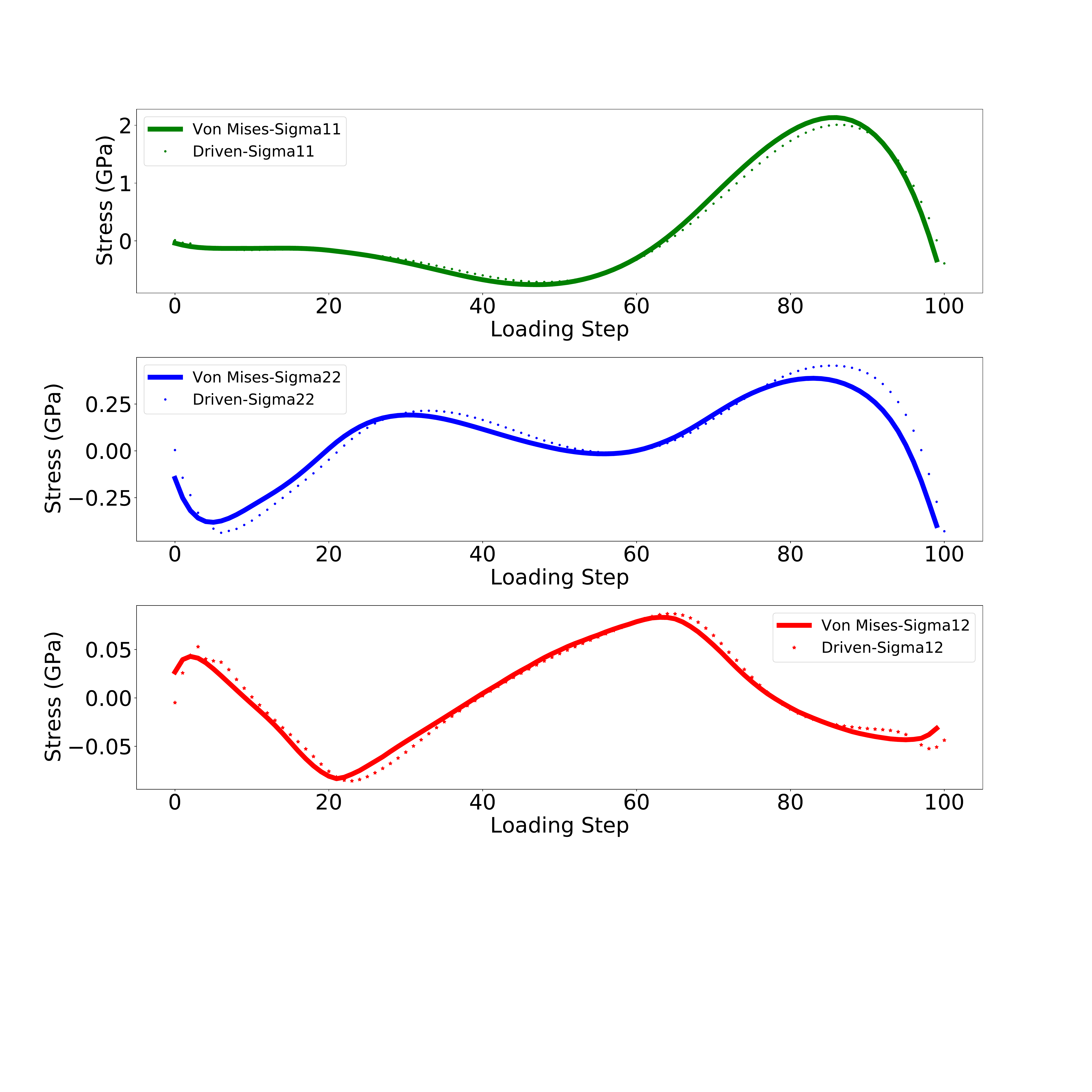}\label{fig-randprob1a}} 
    \caption{Results of the proposed LSTM for capturing path-dependent behaviors of anisotropic microstructures for Part 3 (Section \ref{Sec : Problem3}), (a) Monotonic Loading, (b) Monotonic Loading-unloading, (c) Random Loading-unloading}
    \label{Fig: Test_Problem2}
\end{figure}

It is worth nothing that new loading paths are considered for the testing as well. 
The loading paths are designed to capture various homogenized responses 
via the monotonic loading and unloading and the random loading and unloading conditions.
Figure \ref{Fig: Test_Problem2} depicts (a) monotonic loading, (b) monotonic 
loading-unloading, (c) random loading-unloading to test the capability of deep recurrent 
network in predicting different microstructure. 
The results demonstrate that the path-dependent responses of transversely isotropic microstructures
are captured within 1 percent of error.  

\section{Conclusion}
\label{sec:closure}

This study investigates applicability of the basic long-short term memory (LSTM) network architecture
to capture path-dependent responses of two-dimensional microstructures associated with material heterogeneity and anisotropy.
A single framework of the basic LSTM networks is proposed to learn both elastic and elastoplastic responses
under various loading conditions.
Introducing the averaged history of strain into input enhances inductive biases toward history information of the basic LSTM,
which resolves the lack of mass conservation reported by \citet{hoedt2021mc}.
Applicability of the proposed framework is investigated by two aspects of material responses. 
First, the elasto-plastic constitutive behavior under the plane stress condition is investigated,
where the J2 plasticity yield criterion and isotropic hardening are adopted.
Variation of each material parameter, including elastic properties, yield stress, and hardening modulus,
is considered to account for material heterogeneity associated with the path-dependent responses. 
Second, the homogenized microscopic mechanical response via the finite element analysis is assessed
as a data-driven model for multiscale simulations
Transversely isotropic microstructures are explicitly configured, 
in which heterogeneous anisotropic features are considered 
by changing the pattern and thickness of alternating horizontal layers with elastic and elasto-plastic materials, respectively. 
Three descriptors are adopted to input data to identify anisotropic attributes of each microstructure.
The proposed framework of a single basic LSTM network architecture is examined systematically
associated with various loading and unloading conditions. 
The results of training and testing shows that the proposed data-driven LSTM method 
well captures path-dependent responses at both local constitutive and homogenized microstructural levels. 
The proposed strategy is also proved very effective in capturing the heterogeneous 
and anisotropic responses over wide rages of loading conditions, including monotonic, non-monotonic, 
and random loading-unloading.
Despite the current progress of Deep Learning for capturing complicated material responses, 
less attention has been paid to directly use the basic LSTM networks for path-dependent relationship
between stress and strain tensors for various material heterogeneity and anisotropic under generalized loading conditions,
which is a key component of efficient data-driven multiscale modeling. 
The simplicity and generality of the proposed framework along with the significance of path-dependence, heterogeneity,
and anisotropy in designing structures and materials shows its high potential applicability to various fields. 


\section*{Acknowledgments}
\noindent
This research was supported by Natural Sciences and Engineering Research Council of Canada 
(Discovery Grant, RGPIN-2019-06471). 
The authors thank Prof. Sepp Hochreiter and Dr. Frederik Kratzert for their valuable recommendation and comments.
  

\bibliography{main}

\end{document}